\documentstyle[12pt,aasms4]{article}

\def\be{\begin{equation}}
\def\ee{\end{equation}}
\def\beq{\begin{eqnarray}}
\def\eeq{\end{eqnarray}}
\def\lra#1{\left\langle #1\right\rangle}
\def\part{\partial}
\def\nn{\nonumber}
\def\l{\left}
\def\r{\right}

\def\la{\lambda}
\def\f{{\bf f}}
\def\k{{\bf  k}}

\def\ep{\epsilon}

\def\a{{\bf a}}
\def\A{{\bf A}}
\def\x{{\bf x}}

\def\etal{{\it et al.\ }}
\def\b{{\bf b}}
\def\B{{\bf B}}

\def\v{{\bf v}}
\def\V{{\bf V}}
\def\OV{\overline{\V}}
\def\OB{\overline{\B}}
\def\ob{\overline{B}}

\begin{document}
\baselineskip=24pt
\begin{center}
{\Large\bf The Dependence of Dynamo $\alpha$-Effect on Reynolds
Numbers, Magnetic Prandtl Number, and the Statistics of MHD Turbulence}
\bigbreak
{\large\bf by}
\medbreak
{\large\bf Hongsong Chou} \\
{\it Harvard-Smithsonian Center for Astrophysics, Cambridge, MA 02138,
U.S.A. \\ chou5@fas.harvard.edu}
\end{center}
\begin{abstract}
We generalize the derivation of dynamo coefficient $\alpha$ of Field
\etal(1999) to include the following two aspects: first, the
de-correlation times of velocity field and magnetic field are different;
second, the magnetic Prandtl number can be arbitrary. We find that the
contributions of velocity field and magnetic field to the $\alpha$ effect
are not equal, but affected by their different statistical
properties. In the limit of large kinetic Reynolds number and large
magnetic Reynolds number, $\alpha$-coefficient may not be small if the
de-correlation times of velocity field and magnetic field are shorter
than the eddy turn-over time of the MHD turbulence. We also show that
under certain circumstances, for example if the kinetic helicity and
current helicity are comparable, $\alpha$ depends insensitively on
magnetic Prandtl number, while if either the kinetic helicity or
the current helicity is dominated by the other one, a different magnetic
Prandtl number will significantly change the dynamo $\alpha$ effect.
\end{abstract}


\section{Introduction}
Dynamo theory is an attempt to understand the process of magnetic
field generation by self-inductive action in electrically conducting
fluids. The theory has been used to explain the magnetic field
generation in many celestial objects such as the Sun, the Earth and
the Galaxy. Dynamo theory divides into a kinematic regime and a dynamic
regime. In the kinematic regime, the velocity field is prescribed, and
dynamo theory studies the physics of magnetic field under the determined
velocity field. In general, kinematic dynamo theory focuses solely on the
mathematical solution of the induction equation,
\be
\frac{\part \B}{\part t} = \nabla \times \left ( \V \times \B \right) +
\lambda \nabla^2 \B,
\ee
where $\V$, $\B$ are the velocity and magnetic field, respectively, and
$\lambda$ is the magnetic diffusivity. 

In the dynamic regime, the velocity field can be modified by the magnetic
field, so that the dynamic dynamo theory must then consider both the induction
equation and the momentum equation
\be
\frac{\part \V}{\part t} = -\V \cdot \nabla \V + \B \cdot \nabla \B +
\nu \nabla^2 \V - \nabla P + \f,
\ee
where $\nu$ is the molecular viscosity, $P$ the total pressure, and
$\f$ the external forcing term. The mean-field electrodynamics(MFE)
developed by Steenbeck, Krause and R\"{a}dler(1966) through the
two-scale approach provides essential insights into the relation
between the statistical properties of turbulence and dynamo effects,
namely, the $\alpha$ and $\beta$ parameters of dynamo theory. Let the
large-scale magnetic field and velocity field be $\OB$ and $\OV$, and the
fluctuating components of magnetic field and velocity field be $\b$,
$\v$, respectively. The two-scale separation of MFE gives the equation for
$\OB$ as\footnote{$< \cdot >$ and ${\overline {\mbox{  $\cdot$  }}}$ are
interchangeably used throughout this paper to denote ensemble average.} 
\be
\frac{\part \OB}{\part t} = \nabla \times \left( \OV \times \OB \right)
+ \nabla \times < \v \times \b > + \lambda \nabla^2 \OB.
\ee
The so-called turbulent electromotive force, $\cal{E} = < \v \times \b >$, is
related to the dynamo $\alpha$ and $\beta$ effects through
\be
\cal{E} = \alpha \OB + \beta \nabla \times \OB.
\ee
MFE in the kinematic regime gives (see section 7.3 of Biskamp, 1993,
and references therein)
\be
\alpha = -\frac{\tau}{3} <\v \cdot \nabla \times \v>, \mbox{  }\beta =
\frac{\tau}{3} <\v^2>,
\ee
where $\tau$ is the velocity de-correlation time. Largely as a result of
the development of MFE, the kinematic aspect of dynamo theory has been
broadly understood. Several monographs have been devoted to this
subject(Moffat 1978, Krause and R\"{a}dler 1980, Zeldovich \etal
1983). 

But the nature of dynamo theory in the dynamic regime is still in
debate. The back reaction of the magnetic field on the velocity field
will modify the expressions for the  dynamo $\alpha$ and $\beta$
coefficients. The numerical simulation by Pouquet \etal (1976) was
among the first to point out that with back reaction, to lowest order
in ${\overline B}$ the dynamo $\alpha$ effect should be modified to
\be
\alpha_{br} = \alpha_v + \alpha_b
\ee
where $\alpha_v$ is the $\alpha$ in (5) and $\alpha_b$ is proportional
to $<\b \cdot \nabla \times \b>$\footnote{$\alpha_{br}$ was called {\it
residual torsality} by Pouquet \etal Following Field \etal(1999), we call
$<\b \cdot \nabla \times \b>$ {\it current helicity} in this paper.}.
This new term will reduce the classical kinematic dynamo $\alpha$
effect to a certain degree. Some early criticism of MFE was discussed
by Piddington (1970, 1972abc, 1975ab). He argued that kinematic solar dynamo
theories do not account for the removal of the large amounts of flux
generated each solar cycle. Recent objections to dynamo action have
their root in the problem of small-scale magnetic fields. For
astrophysical systems such as the Galaxy, where magnetic Reynolds
number $R_m = Lv_0/\lambda$ is large, some authors(Cattaneo and
Vainshtein 1991, Vainshtein and Cattaneo 1992, Gruzinov and Diamond
1996) argue that the magnetic energy at small scales, $<\b^2>$, is much
greater than the magnetic energy at large scales, $\OB^2$, through the
relation
\be
<\b^2> = R_m \OB^2.
\ee
According to Cattaneo and Vainshtein (1991), the magnetohydrodynamic
turbulent dynamo will stop operating as soon as relation (7) is
obtained, a process that can happen in much shorter time than turbulent
eddy turn-over time (Kulsrud and Anderson, 1992). Gruzinov and Diamond
(1994) base their argument on the conservation of squared vector
potential for 2D MHD and magnetic helicity for 3D MHD, and claim that
dynamo $\alpha$ effect will be quenched in systems of high magnetic Reynolds
number as follows
\be
\alpha = \frac{\alpha_v}{1 + R_m {\left( \OB/v_0 \right)}^2 }
\ee
where $\alpha_v = -\tau/3<\v \cdot \nabla \times \v>$ is the classical
result in (5). The numerical simulation with periodic boundary
conditions by Cattaneo and Hughes (1996) supports relation (8) for the
particular value $R_m=100$ and various values of ${\overline
B}$. Relation (8) is completely different from a previous estimate of
$\alpha$ effect made by Kraichnan (1979). He argued that even in the
high magnetic Reynolds number limit, the $\alpha$ effect will not be
quenched. Rather, it has the following relation with the classical
estimate, $\alpha_v$, and $\overline B$,
\be
\alpha \sim \frac{\alpha_v}{1+{\left( {\overline B}/v_0 \right)}^2}.
\ee

Kulsrud (1999) questioned the derivations of Gruzinov and Diamond
(1994, 1995, 1996) by arguing that one of the their results, $\alpha =
\beta \OB \cdot \nabla \times \OB$, for large $R_m$, leaves out the
contribution from $\v \cdot \nabla \times \v$ completely. Note that
Gruzinov and Diamond derived (8) from the conservation of magnetic
helicity, $\A \cdot \B$, which is based solely on the induction
equation. In the dynamic regime, a velocity field that is constantly driven
by external force will be modified by the growing magnetic
field, a process that cannot be wholly understood by using only the
conservation of magnetic helicity. Leaving this process out of the
discussion of dynamo action is questionable, and a more complete
account of dynamo theory in the dynamic regime must also consider
momentum equation and the role of external forcing terms in maintaining the
turbulence. Field, Blackman and Chou (1999) considered a simplified
model of MHD turbulence. The external forcing term in their model has
the freedom to drive a MHD turbulence that is independent of the
presence of any large-scale magnetic field. The turbulent velocity field
and magnetic field were treated on an equal footing. Their result on the
$\alpha$ dynamo effect depends on the statistical properties of MHD
turbulence that are independent of $\OB$, and agrees approximately with
that of (9) but disagrees with (8).   

In this paper, we generalize the work by Field, Blackman and Chou
(1999) in the following two aspects: First, we relax the assumption
made by Field, Blackman and Chou that the de-correlation times of velocity
field and magnetic field are the same. Second, we relax the assumption
in Field \etal that magnetic Prandtl number, $Pr = \nu/\la$, is one. We
also give brief discussions on how the nonlinear interaction between
$\OB$-induced quantities ($\v^{\prime}$ and $\b^{\prime}$ in Field
\etal) changes the final result for $\alpha$, and compare our result
with (8) and discuss where the difference originates. This work can be
considered as a generalization of Field, Blackman and Chou (1999), yet
itself is complete and self-contained. To simplify matters, we
discuss only the dynamo $\alpha$ effect (not $\beta$ effect) in this work by
assuming a constant large-scale magnetic field, $\OB$.

The structure of this paper is as follows: in section 2, we discuss our
model based on the work of Field \etal In section 3 and appendix A, we
give our derivation of the coefficient of dynamo $\alpha$ effect. In
section 4, we talk about the dependence of the $\alpha$ effect on the
kinetic Reynolds number $R_e$, the magnetic Reynolds number $R_m$ and
the magnetic Prandtl number $Pr(=\nu/\la)$ based on our derivation in section
4. In section 5, we discuss possible modifications to our result in the
presence of strong nonlinear interaction between $\OB$-dependent
components of the MHD turbulence, and compare our result with previous
derivations of the $\alpha$ effect. Conclusions are made in the final
section, section 6.

In Table 1, we list the notations used throughout this paper. The
physical meanings of these quantities are also explained briefly in
this table. 

\section{Separation of ${\v}^{\prime}$, ${\b}^{\prime}$ from ${\v}^{(0)}$
and ${\b}^{(0)}$}
As in the work of Field, Blackman and Chou(1999), we consider only
incompressible fluids. In our model, we first distinguish four
different scales in the system. We denote the size of turbulence energy
containing eddy as $l_0$, and denote the dissipation scale of the
turbulence as $l_D$. For turbulence of large kinetic Reynolds number, we have
$l_0 \gg l_D$. An ``ensemble average scale'', denoted by $L \gg l_0$, is
used to carry out the calculations of averaged quantities, $< \cdot
>$. Finally, we denote the scale of the whole physical system as $S$,
which is the typical scale for the variations of averaged quantities $<
\cdot >$. So we have the relation $S \gg L \gg l_0 \gg l_D$. In our model,
there are two large-scale quantities, $\OB$ and $\OV$. We assume both
of these  quantities are constant, i.e., $S \rightarrow \infty$. We
set our reference frame to that moving at $\OV$ and henceforth omit
terms of $\OV$. The presence of non-zero $\OB$ is related to the $\alpha$
effect in the following two aspects: first, $\OB$ can be amplified due
to the dynamo $\alpha$ effect; second, for large $\OB$, $\alpha$ effect
will be quenched.  

Next, we make the distinction between the $\OB$-independent components,
${\v}^{(0)}$ and ${\b}^{(0)}$ from the $\OB$-induced components,
${\v}^{\prime}$ and ${\b}^{\prime}$. In the absence of a large-scale
magnetic field $\OB$, we assume that a homogeneous, isotropic,
steady-state MHD turbulence be maintained by an external force, which
operates at scale $\sim l_0$. To simplify matters, we consider only the
case that the external force is homogeneous, isotropic and helical. We
further assume that within the scale range, $[\epsilon l_0,
{\epsilon}^{-1} l_0]$ where $\epsilon > 1$, the turbulence is largely
hydrodynamic in character. This assumption can be justified if the
external force, which is independent of the velocity and the magnetic
fields, is the only external energy source to drive the velocity
field. A homogeneous, isotropic, helical external force will constantly
drive a homogeneous, isotropic, helical velocity field near the forcing
scale. The equipartition between the velocity field and magnetic field
is not achieved within range $[\epsilon l_0, {\epsilon}^{-1}
l_0]$. Rather, the kinetic energy within this range surpasses the magnetic
energy, a phenomenon that has been found in many numerical simulations
(see below). From the scale ${\epsilon}^{-1}l_0$ down to viscous
cut-off scale $\sim l_D$ of the velocity field, equipartition can be maintained
according to the MHD turbulence model of Kraichnan(1965) or Goldreich
and Sridhar(1995). The simulation by Pouquet \etal(1976) with simplified
DIA equations shows that $\epsilon \sim 3$. Simulations by Kida
\etal(1991) with a spectral method gives $\epsilon \sim 5$. Both
authors obtained homogeneous, isotropic MHD turbulence with no
large-scale magnetic field present, i.e., $\OB=0$. We denote the
velocity and magnetic field in this kind of homogeneous, isotropic
turbulence as ${\v}^{(0)}$ and ${\b}^{(0)}$, and clearly they satisfy
the equation
\be
\part_t {\b}^{(0)} = \nabla \times \left( {\v}^{(0)} \times {\b}^{(0)}
\right) + \la {\nabla}^2 {\b}^{(0)},
\ee
and
\be
\part_t {\v}^{(0)} = -{\v}^{(0)} \cdot \nabla {\v}^{(0)} + {\b}^{(0)}
\cdot \nabla {\b}^{(0)} + \nu {\nabla}^2 {\v}^{(0)} - \nabla P^{(0)} + \f,
\ee
where $\la$ and $\nu$ are magnetic resistivity and molecular viscosity,
respectively. $\f$ is the external forcing term. 

As we mentioned before, $\OB$ can be amplified by $\alpha$ effect, and
a growing $\OB$ will also attempt to reduce $\alpha$ dynamo
effect. To tackle this problem, we impose a large-scale magnetic field,
$\OB$, on the MHD turbulence of ${\v}^{(0)}$ and ${\b}^{(0)}$, and then
follow the dynamics of the MHD turbulence system. ${\v}^{(0)}$ will
stretch $\OB$, so that a new component, ${\b}^{\prime}$, of magnetic
field that is dependent on $\OB$, is generated. The new magnetic field,
$\OB + {\b}^{\prime} + {\b}^{(0)}$, will exert Lorentz force on the
velocity field so that a new component of the velocity field,
${\v}^{\prime}$, which depends on $\OB$, is in turn generated.
${\v}^{(0)}$ and ${\b}^{(0)}$ still satisfy equations (10) and (11) even
in the presence of $\OB$. This is because the change in the total
velocity field, $\v^{\prime} = \v - \v^{(0)}$, is due to $\OB$, while
the component $\v^{(0)}$ is driven primarily by external forcing term
$\f$, which is independent of $\OB$. Bearing this in mind, and
assuming that we have ${\v}^{(0)}$ and ${\b}^{(0)}$ already at our
disposal\footnote{By, say, numerical simulations of equations (10) and
(11), or observations of MHD turbulence with no large-scale magnetic
field present.}, we write down the relation between the known
quantities, \{${\v}^{(0)}$, ${\b}^{(0)}$, $\OB$\}, and the unknown
quantities, \{${\v}^{\prime}$, ${\b}^{\prime}$\}, as
\be
\part_t {\v}^{\prime} = \OB \cdot \nabla \left( {\b}^{\prime} +
{\b}^{(0)} \right) + \nu {\nabla}^2 {\v}^{\prime},
\ee
\be
\part_t {\b}^{\prime} = \OB \cdot \nabla \left( {\v}^{\prime} +
{\v}^{(0)} \right) + \lambda {\nabla}^2 {\b}^{\prime}.
\ee
(12) and (13) show that the generation of $\OB$-dependent component of
magnetic field is due to the stretching of $\OB$ by both the newly
generated, $\OB$-dependent ${\v}^{\prime}$, and the $\OB$-independent
velocity component $\v^{(0)}$. The generation of $\OB$-dependent
component of velocity field is due to the Lorentz force that is exerted
by the combination of ${\b}^{\prime}$ and ${\b}^{(0)}$. In this sense,
the terms involving $\v^{(0)}$ and $\b^{(0)}$ in (12) and (13) can be
considered as sources for $\v^{\prime}$ and $\b^{\prime}$, thus the
nonlinearity of ${\v}^{\prime}$ and ${\b}^{\prime}$ will largely come from the
nonlinearity of the MHD turbulence, namely, ${\v}^{(0)}$ and
${\b}^{(0)}$. Therefore, we expect that the statistical properties of
${\v}^{\prime}$ and ${\b}^{\prime}$ depend on the statistical
properties of ${\v}^{(0)}$ and ${\b}^{(0)}$.
 
In deriving equations (12) and (13), we made an important
assumption. The validity of this assumption has been discussed in
Field, Blackman and Chou (1999). The assumption is that the ratio of
the decorrelation time $\tau_{dcor}$ of both magnetic and velocity fields,
to the eddy turn-over time, $\tau_{eddy}$, is small: 
\be
\frac{\tau_{dcor}}{\tau_{eddy}} < 1.
\ee
With such assumption, we dropped the following nonlinear terms: ${\b}^{\prime}
\cdot \nabla {\b}^{\prime}$, ${\b}^{\prime} \cdot \nabla {\b}^{(0)}$,
${\b}^{\prime} \cdot \nabla {\v}^{\prime}$, ${\b}^{\prime} \cdot
\nabla {\v}^{(0)}$, ${\v}^{\prime} \cdot \nabla {\b}^{\prime}$,
${\v}^{\prime} \cdot \nabla {\b}^{(0)}$, ${\v}^{\prime} \cdot \nabla
{\v}^{\prime}$, ${\v}^{\prime} \cdot \nabla {\v}^{(0)}$, ${\b}^{(0)}
\cdot \nabla {\b}^{\prime}$, ${\b}^{(0)} \cdot \nabla {\v}^{\prime}$,
${\v}^{(0)} \cdot \nabla {\v}^{\prime}$, and ${\v}^{(0)} \cdot \nabla
{\b}^{\prime}$. These terms are all smaller than the time derivative
terms, $\part_t \b^{\prime}$ and $\part_t \v^{\prime}$. In fact, the
ratio of these two terms to the nonlinear terms that we dropped is
approximately $\frac{\tau_{eddy}}{\tau_{dcor}}$. We make this assumption
because (for more details, see also Field \etal, 1999) numerical
simulations by several authors (Pouquet \etal 1976, Kida \etal 1991,
Brandenberg 2000) have shown that the MHD turbulence within range
$[\epsilon l_0, {\epsilon}^{-1} l_0]$ for $\epsilon > 3$, is dominated
by velocity field. In other words, the MHD turbulence is largely hydrodynamic
within this range, where most of the energy is concentrated. For
pure hydrodynamic turbulence, experiments (Pope 1994) have shown
that\footnote{Our recent numerical simulation(Chou and Field, 2000) of
3D MHD turbulence also supports the validity of assumption (14). We
found that for 3D MHD turbulence in steady state ($50 < R_m < 300$,
${\overline B} \leq v_{rms}$), $\tau_{dcor}/\tau_{eddy} \sim 0.22$ for
velocity field and $\sim 0.25$ for magnetic field. Similar results are
also obtained by Brandenberg(private communication).}
\be
\frac{\tau_{dcor}}{\tau_{eddy}} = 0.2 - 0.3.
\ee
Therefore the time derivative terms in (12) and (13) are greater than the
nonlinear terms that we dropped to get (12) and (13). Note that the
assumption of (14) does not imply that the turbulence is of only one
scale $l_0$. In fact, the turbulence in this work can be of multiple
scales. However, if the turbulence at scales $\leq {\epsilon}^{-1}
l_0$ is in approximate equipartition, its contribution to the
electromotive force, $<\v \times \b>$, will be small because of the
Alfv\'{e}nic wave-like motions of the MHD turbulence in the
equipartition scales. In other words, we believe that the dynamo
$\alpha-$effect to be discussed in the next few sections is mostly
attributed to the turbulence within scales $[\epsilon l_0,
{\epsilon}^{-1} l_0]$ for $\epsilon > 3$, and various numerical
simulations mentioned above support such picture.

In this work, the ratio of ${\overline B}$ to the rms of turbulent velocity, $v_{rms}$, can
be of any value; therefore we keep all terms that involve $\OB$ in the
MHD equations for $\v^{\prime}$ and $\b^{\prime}$. Indeed, by keeping
all terms of any order in ${\overline B}/v_{rms}$, we include the
nonlinearity from the back reaction of $\OB$ to the turbulence. Also,
under condition (14), our treatment of $\alpha$ effect does not require
any constraint on the ratio of $\overline B$ to the rms of magnetic
field, $b_{rms}$. Thus the back reaction of $\b$ on the turbulence is
also taken into account in our model. This is different from classical
dynamo theory that assumes that $b_{rms}$ is smaller than $\overline B$. 

In the following, we assume that equations (10) and (11) have been solved
so we have ${\v}^{(0)}$ and ${\b}^{(0)}$ at our disposal. In fact, we
require only the statistical properties of these zeroth order velocity
and magnetic fields. Now, equipped with equations (12) and (13), and
the known quantities ${\v}^{(0)}$ and ${\b}^{(0)}$, we will be able to
calculate the relation between $\{{\v}^{\prime}$, ${\b}^{\prime}\}$
and $\{{\v}^{(0)}, {\b}^{(0)}\}$ and, furthermore, calculate the
electromotive force, $<\v \times \b>$ and relate it to dynamo
$\alpha$-effect. The calculation is performed in next section and
appendix A.

\section{The Derivation of Dynamo $\alpha$-coefficient In the Presence
of Nonlinear Effects of ${\v}^{(0)}$ and ${\b}^{(0)}$}
Throughout this paper, we define the Fourier transform ${\hat{F}}$ of a
function $F$ as
\be
\hat{F}(\k,\omega)={\left(\frac{1}{2\pi}\right)}^4\int_V \!
\int_{-\infty}^{+\infty}F(\x,t)e^{-i(\k\cdot\x-\omega t)} dt d\x;
\ee
i.e., we transform $F(\x,t)$ both in time and in space. This method is
similar to the one used by Roberts and Soward(1975), and we assume
$\OB = constant$ throughout this work. The equations of ${\b}^{\prime}$
and ${\v}^{\prime}$, (12) and (13), are transformed into Fourier space as
\be
\left( -i\omega + \nu k^2 \right) {\hat{v}}^{\prime}_j = i(\k \cdot
\OB)({\hat{b}}^{\prime}_j + {\hat{b}}^{(0)}_j),
\ee
\be
\left( -i\omega + \la k^2 \right) {\hat{b}}^{\prime}_j = i(\k \cdot
\OB)({\hat{v}}^{\prime}_j + {\hat{v}}^{(0)}_j).
\ee
Here $j = 1,2,3$. We will use the Einstein summation convention throughout
this paper. Solving (17) and (18) gives the following relations:
\be
{\hat{b}}^{\prime}_j = \xi_{\la}(\k,\omega) {\hat{u}}^{(0)}_j -
\zeta(\k,\omega) {\hat{b}}^{(0)}_j,
\ee
\be
{\hat{v}}^{\prime}_j = \xi_{\nu}(\k,\omega) {\hat{b}}^{(0)}_j -
\zeta(\k,\omega) {\hat{u}}^{(0)}_j,
\ee
where
\be
\xi_{\la}(\k,\omega) = \frac {\frac {i(\k \cdot \OB)}{-i\omega + \la
k^2}} {1+\frac{(\k \cdot \OB)^2}{(-i\omega+\la k^2)(-i\omega+\nu k^2)}},
\ee
\be
\xi_{\nu}(\k,\omega) = \frac {\frac {i(\k \cdot \OB)}{-i\omega + \nu
k^2}} {1+\frac{(\k \cdot \OB)^2}{(-i\omega+\la k^2)(-i\omega+\nu k^2)}},
\ee
and
\be
\zeta (\k,\omega) = \frac {\frac {(\k \cdot \OB)^2}{(-i\omega+\la
k^2)(-i\omega+\nu k^2)}}{1+\frac{(\k \cdot \OB)^2}{(-i\omega+\la
k^2)(-i\omega+\nu k^2)}}.
\ee
The solutions (19) and (20) relate the induced velocity field and
magnetic field to the background $\b^{(0)}$ and $\v^{(0)}$. Except for
the condition (14), we set no constraint on these zeroth order
components, i.e., the components that are independent of $\OB$. 

To calculate dynamo $\alpha$-effect, we use the following relation
\be
\lra{\v \times \b}_n = \alpha \ob_n.
\ee
Here $\v = \v^{(0)} + \v^{\prime}$, and $\b = \b^{(0)} +
\b^{\prime}$. We then apply the properties of the Fourier transform and
write the correlation between $\v$ and $\b$ as 
\be
\lra{v_l(\x,t)b_j(\x,t)}=\int \! \int d{\k} \, d{\omega} \, e^{i(\k \cdot \x
- \omega t)} {\left[\int \! \int d{\k}^{\prime} d{\omega}^{\prime}
\lra{{\hat{v}}_l({\k}^{\prime},{\omega}^{\prime}){\hat{b}}_j(\k-{\k}^{\prime},
\omega-{\omega}^{\prime})}\right]}.
\ee
Let $g_{lj}=\lra{{\hat{v}}_l({\k}^{\prime},{\omega}^{\prime}){\hat{b}}_j(\k-{\k}^
{\prime},\omega-{\omega}^{\prime})}$; then with (19) and (20), we have
\beq
g_{lj} &=& \left[ 1-\zeta(\k - {\k}^{\prime}, \omega - 
{\omega}^{\prime})-\zeta({\k}^{\prime}, {\omega}^{\prime}) \r. \nn\\
&+& \l.
\zeta({\k}^{\prime},{\omega}^{\prime})\zeta(\k - {\k}^{\prime}, \omega
- {\omega}^{\prime})\right]{\hat{P}}^{00}_{lj}(\k - {\k}^{\prime},
\omega - {\omega}^{\prime})\delta(\k,\omega) \nn\\
&+& \left[\xi_{\la}(\k - {\k}^{\prime}, \omega - {\omega}^{\prime}) -
\zeta({\k}^{\prime},{\omega}^{\prime})\xi_{\la}(\k - {\k}^{\prime},
\omega - {\omega}^{\prime})\right]{\hat{Q}}^{00}_{lj}(\k - {\k}^{\prime},
\omega - {\omega}^{\prime})\delta(\k,\omega) \nn\\
&+& \left[\xi_{\nu}({\k}^{\prime}, {\omega}^{\prime}) -
\xi_{\la}({\k}^{\prime},{\omega}^{\prime})\zeta(\k - {\k}^{\prime},
\omega - {\omega}^{\prime})\right]{\hat{S}}^{00}_{lj}(\k - {\k}^{\prime},
\omega - {\omega}^{\prime})\delta(\k,\omega)  \nn\\
&+& \left[\xi_{\nu}({\k}^{\prime}, {\omega}^{\prime})\xi_{\la}(\k -
{\k}^{\prime}, \omega - {\omega}^{\prime})\right]{\hat{P}}^{00}_{jl}({\k}^{\prime},{\omega}^{\prime})\delta(\k,\omega).
\eeq
Here we've used the homogeneity and isotropy of $\v^{(0)}$ and $\b^{(0)}$, and the
following definitions of correlation functions and their corresponding Fourier
transforms:
\be
\lra{v_l^{(0)}(\x,t)v_j^{(0)}(\x^{\prime},t^{\prime})}\equiv
Q_{lj}^{00}({\x}^{\prime}-\x,t^{\prime}-t),
\ee
\be
\lra{v_l^{(0)}(\x,t)b_j^{(0)}(\x^{\prime},t^{\prime})}\equiv
P_{lj}^{00}({\x}^{\prime}-\x,t^{\prime}-t),
\ee
\be
\lra{b_l^{(0)}(\x,t)b_j^{(0)}(\x^{\prime},t^{\prime})}\equiv
S_{lj}^{00}({\x}^{\prime}-\x,t^{\prime}-t),
\ee
\be
\lra{{\hat{v}}_l^{(0)}(\k,\omega){\hat{v}}_j^{(0)}(\k^{\prime},{\omega}^{\prime})}=
{\hat{Q}}_{lj}^{00}({\k}^{\prime},{\omega}^{\prime})\delta(\k+{\k}^{\prime},\omega+{\omega}^{\prime}),
\ee
\be
\lra{{\hat{v}}_l^{(0)}(\k,\omega){\hat{b}}_j^{(0)}(\k^{\prime},{\omega}^{\prime})}=
{\hat{P}}_{lj}^{00}({\k}^{\prime},{\omega}^{\prime})\delta(\k+{\k}^{\prime},\omega+{\omega}^{\prime}),
\ee
\be
\lra{{\hat{b}}_l^{(0)}(\k,\omega){\hat{b}}_j^{(0)}(\k^{\prime},{\omega}^{\prime})}=
{\hat{S}}_{lj}^{00}({\k}^{\prime},{\omega}^{\prime})\delta(\k+{\k}^{\prime},\omega+{\omega}^{\prime}).
\ee
With such definitions of correlation functions and the properties of
Fourier transform, (24) can be further calculated as (see
Appendix A for details)
\be
\lra{\v \times \b}_n = M_n + N_n,
\ee
where
\be
M_n=\int \! \int d\k d\omega
\left[\xi_{\la}(\k,\omega)-\zeta(-\k,-\omega)\xi_{\la}(\k,\omega)
\right]\ep_{nlj}{\hat{Q}}^{00}_{lj}(\k,\omega),
\ee
and
\be
N_n=\int \! \int d\k d\omega
\left[\xi_{\nu}(-\k,-\omega)-\zeta(\k,\omega)\xi_{\nu}(-\k,-\omega)
\right]\ep_{nlj}{\hat{S}}^{00}_{lj}(\k,\omega).
\ee
The above integrals can be calculated with the following assumptions
about the MHD turbulence. Suppose that we can separate variables in the
Fourier transformed correlation functions and write 
${\hat{Q}}^{00}_{lj}(\k,\omega)={\hat{X}}^{00}_{lj}(\k){\hat{T}}(\omega)$
and ${\hat{S}}^{00}_{lj}(\k,\omega)={\hat{O}}^{00}_{lj}(\k){\hat{J}}(\omega)$.
In this paper we also assume that $T(\varphi)=e^{-|\varphi|{\omega}_k}$
and $J(\varphi)=e^{-|\varphi|{\chi}_k}$ where $\varphi=t-t^{\prime}$,
$\omega_k$ and $\chi_k$ are positive and functions of the wave vector, and
$\omega_k = 1/{\tau_{v\_cor}}$ and $\chi_k=1/{\tau_{b\_cor}}$ where
$\tau_{v\_cor}$ is the de-correlation time of velocity field at $\k$,
and $\tau_{b\_cor}$ is the de-correlation time of magnetic field at
$\k$. Again, because of the homogeneity and isotropy, we have
that (Krause and R\"adler, 1980, p. 75) 
\be 
{\hat{X}}^{00}_{lj}(\k)=A(k)\delta_{lj}+B(k) k_l k_j +
i\ep_{klj}k_kC(k),
\ee
and
\be
{\hat{O}}^{00}_{lj}(\k)=D(k)\delta_{lj}+G(k) k_l k_j +
i\ep_{klj}k_kH(k),
\ee
where $A,B,C,D,G,H$ are functions of $k=|\k|$. Here we use $C(k)$ and
$H(k)$ to denote the helical parts of the spectra of the velocity field
and magnetic field, respectively. With these assumptions
and tedious evaluations (see Appendix A for details), we have much
simpler integral forms of $M$ and $N$ as
\be
M_n=-\frac{1}{2} \int^{+\infty}_0 dk \left
[ \frac{2\la}{\la+\nu}\Delta_v(k)\Pi_v(k)\right]{\ob}_n,
\ee
\be
N_n=+\frac{1}{2} \int^{+\infty}_0 dk \left
[ \frac{2\nu}{\nu+\la}\Delta_b(k)\Pi_b(k)\right]{\ob}_n,
\ee
where
\be
\Delta_v(k)\equiv \frac{8\pi k^4C(k)}{\omega_k},
\ee
\be
\Delta_b(k)\equiv \frac{8\pi k^4H(k)}{\chi_k},
\ee
\beq
\Pi_v(k) &=& \left[{\left(\frac{\nu k^2}{|\OB|k}\right)}^2 - \frac{\nu^2
k^4 \sqrt{\la\nu
k^4}}{(|\OB|k)^3}\arctan\left(\frac{|\OB|k}{\sqrt{\la\nu k^4}}\right)
\right] \nn\\ &+& \frac{\omega_k-\nu k^2}{\omega_k+\nu k^2} \left
[ {\left( \frac{\omega_k+\nu k^2}{|\OB|k}\right) }^2 - \frac{(\omega_k+\nu
k^2)^2\sqrt{\omega_k^2+(\la+\nu)k^2\omega_k+\la\nu
k^4}}{(|\OB|k)^3} \times \r. \nn\\ && \l. \arctan{\left(\frac{|\OB|k}{\sqrt{\omega_k^2+(\la+\nu)k^2\omega_k+\la\nu
k^4}}\right)}\right],
\eeq
\beq
\Pi_b(k) &=& \left[{\left(\frac{\la k^2}{|\OB|k}\right)}^2 - \frac{\la^2
k^4 \sqrt{\nu\la
k^4}}{(|\OB|k)^3}\arctan\left(\frac{|\OB|k}{\sqrt{\nu\la k^4}}\right)
\right] \nn\\ &+& \frac{\chi_k-\la k^2}{\chi_k+\la k^2} \left
[ {\left( \frac{\chi_k+\la k^2}{|\OB|k}\right) }^2 - \frac{(\chi_k+\la
k^2)^2\sqrt{\chi_k^2+(\nu+\la)k^2\chi_k+\nu\la
k^4}}{(|\OB|k)^3} \times \r. \nn\\ && \l. \arctan{\left(\frac{|\OB|k}{\sqrt{\chi_k^2+(\nu+\la)k^2\chi_k+\nu\la
k^4}}\right)}\right].
\eeq
With (24), (33), (38) and (39), we finally have an expression for $\alpha$
\be
\alpha = -\frac{1}{2} \int_0^{+\infty}dk
\left[\frac{2\la}{\la+\nu}\Delta_v(k)\Pi_v(k) -
\frac{2\nu}{\nu+\la}\Delta_b(k)\Pi_b(k) \right].
\ee

\section{The Dependence of $\alpha$ on Kinetic Reynolds Number $R_e$,
Magnetic Reynolds Number $R_m$ and Prandtl Number $Pr$}
Before we discuss the dependence of $\alpha$ on different dimensionless
parameters of the turbulence, we explain the physical meanings of the
notations introduced in above derivation. $C(k)$ in relation (40) was
first introduced in the correlation spectrum of zeroth order velocity
field, i.e., equation (36). It is the helical part of the correlation
function of velocity field $\v^{(0)}$. In fact, one can easily obtain
the following relation
\be
\int 8\pi k^4C(k) dk = < \v^{(0)} \cdot \nabla \times \v^{(0)} >,
\ee
where $< \cdot >$ denotes spatial average. This shows that $8\pi
k^4C(k)$ can be regarded as the spectrum of kinetic helicity. In
parallel, we may regard $8\pi k^4H(k)$ as the spectrum of the current
helicity of magnetic field according to the following relation  
\be
\int 8\pi k^4H(k) dk = < \b^{(0)} \cdot \nabla \times \b^{(0)} >.
\ee
$1/{\omega_k}$ is the de-correlation time of $\v^{(0)}$ at scale
$\sim 1/k$, and $1/{\chi_k}$ is the de-correlation time of $\b^{(0)}$ at
scale $\sim 1/k$. These two parameters are statistical
properties of MHD turbulence of $\v^{(0)}$ and $\b^{(0)}$. They are
independent of $\OB$ but certainly affected by the external forcing
term, $\f$, as $\v^{(0)}$ and $\b^{(0)}$ are driven by this
force. We believe that these two parameters represent the effect of turbulent
nonlinear interactions between different modes of $\v^{(0)}$ and
$\b^{(0)}$, such as the nonlinear cascade of energy from the forcing
scale $\sim l_0$ down to dissipation scales. Note that in our model,
$\omega_k$ and $\chi_k$ are functions of the magnitude of wave vector
$\k$, not its direction. This is because we made the assumption that
the MHD turbulence of $\v^{(0)}$ and $\b^{(0)}$ is isotropic. This is
certainly a simplification. Because the turbulence is driven by
external force $\f$, if $\f$ is not isotropic, anisotropic structures
of $\v^{(0)}$ and $\b^{(0)}$ will develop in turbulence, and the
$\alpha$-effect cannot be represented by a parameter as shown in
(24). Rather, we must write $\alpha$ as a tensor. If external force is
not strongly anisotropic, equation (44) may be used to approximate the
isotropic component of the $\alpha$-effect. For strongly anisotropic
turbulence, the $\alpha$ calculated in above section only gives the
$\alpha$-dynamo effect from the isotropic parts of the velocity and
magnetic field, which may play much less important role than the
anisotropic parts.\footnote{Another concern of such isotropy
assumption, is that recent work on MHD turbulence (see Moran 2000 and
references therein) shows that anisotropic small-scale structure of
$\v^{(0)}$ and $\b^{(0)}$ develops within the inertial range; therefore
the direct cascade of both kinetic and magnetic energy will not be
isotropic. The numerical simulation by Moran(2000) shows that, if MHD
turbulence at the forcing scale is isotropic, small anisotropic
structure only develops at scales five or six times smaller than the
forcing scale. That is, for the MHD turbulence of $\v^{(0)}$ and
$\b^{(0)}$ within the scale range $[\epsilon l_0, {\epsilon}^{-1} l_0]$
for $\epsilon \sim 5$, $\v^{(0)}$ and $\b^{(0)}$ are still isotropic to a
large degree. But for scale $l \ll \epsilon^{-1} l_0$, because two
initially adjacent magnetic field lines can stay correlated after they
wander a distance of $l$, small-scale anisotropic structures will
develop according to the anisotropic energy cascade mechanism (see
Goldreich and Sridhar, 1995). However, one must notice that such anisotropy is
defined with respect to a magnetic field $\b^{(0)}(l^{\prime})$ for
certain scale $l^{\prime} > l$ at certain
instance. $\b^{(0)}(l^{\prime})$ itself is isotropic when averaged over
several turbulence outer scale $l_0$, or when averaged over many eddy
turn-over time of the turbulence. Because the $\alpha$ effect considered
in this work is an ensemble averaged quantity over the scale $L \gg
l_0$, we believe the anisotropy of MHD turbulence at very small scales
will not change our final results qualitatively.}   

If we consider the special case that molecular viscosity is the same as
magnetic resistivity, $\la = \nu$, we have the following expression
\be
\alpha = -\frac{1}{2} \int^{+\infty}_0 dk [\tau_{v,k} 8\pi k^2C(k) \Pi_v(k) -
\tau_{b,k} 8\pi k^2H(k) \Pi_b(k)],
\ee
where
\be
\tau_{v,k} = \frac{1}{\omega_k}, \tau_{b,k} = \frac{1}{\chi_k}, 
\ee
For fixed, negligible but non-zero, values of $\nu=\lambda$ and finite
values of $\tau_{v,k}$ and $\tau_{b,k}$, in the limit of ${\overline B}
\rightarrow 0$, because
\be
\arctan(\xi) = \sum_{k=0}^{\infty}\frac{(-1)^{k}
{\xi}^{2k+1}}{2k+1}\mbox{, for $\xi \rightarrow 0$,}
\ee
we have
\be
\Pi_{v(b)} \stackrel{{\overline B} \rightarrow 0}{\longrightarrow} \frac{1}{3}.
\ee
With (38) and (39), we have the $\alpha$ effect in the limit of small
${\overline B}$ as
\be
\alpha \approx -\frac{1}{3}\left(\tau_v <\v^{(0)} \cdot \nabla \times
\v^{(0)}> - \tau_b <\b^{(0)}
\cdot \nabla \times \b^{(0)}> \right).
\ee
In the limit of $\tau_v = \tau_b$, the same result has been obtained by
Field \etal(1999). Here $\tau_v$ and $\tau_b$ are the typical
de-correlation times of velocity field and magnetic field,
respectively. Note that the current helicity term, $\frac{1}{3} \tau_b
<\b^{(0)} \cdot \nabla \times \b^{(0)}>$, cannot be neglected. It
appears naturally when $\v$ and $\b$ are treated on an equal footing,
which was not considered by kinematic dynamo theory.

We expect that $\alpha$-effect is quenched if $\OB$ is moderate or
large, i.e., ${\overline B} \geq v_{0}$ where $v_{0}$ is the
turbulent velocity at scale $l_0$. This can be seen from (47) under the
conditions that: (a)${\overline B} \geq v_{0}$, (b)$\nu (= \lambda)$
is small, and (c)the de-correlation time is smaller than the eddy turn-over
time but larger than the Alfv\'{e}n time $\sim 1/k{\overline B}$.  Because
\be
\arctan(\xi) \rightarrow \frac{\pi}{2} \mbox{, for $\xi \rightarrow
\infty $,}
\ee
with conditions (a) and (b), we have
\be
{\left(\frac{\nu k^2}{|\OB|k}\right)}^2 - \frac{\nu^2
k^4 \sqrt{\la\nu
k^4}}{(|\OB|k)^3}\arctan\left(\frac{|\OB|k}{\sqrt{\la\nu k^4}}\right)
\longrightarrow {\left(\frac{\nu k^2}{|\OB|k}\right)}^2
\ee
for small $\nu, \la$ and moderate or large ${\overline B}$. Condition
(c) requires that: (d) ${\omega_k} / {{\overline B}k} < 1$. Thus with
(d), (a) and (b), we have that up to the second order in ${\omega_k} /
{{\overline B}k}$
\beq
&& {\left( \frac{\omega_k+\nu k^2}{|\OB|k}\right) }^2 - \frac{(\omega_k+\nu
k^2)^2\sqrt{\omega_k^2+(\la+\nu)k^2\omega_k+\la\nu k^4}}{(|\OB|k)^3}
\times  \nn\\ && \arctan{\left(\frac{|\OB|k}{\sqrt{\omega_k^2+(\la+\nu)k^2\omega_k+\la\nu
k^4}}\right)} \longrightarrow {\left( \frac{\omega_k}{|\OB|k}\right) }^2.
\eeq
Also, to the second order in ${\chi_k} / {{\overline B}k}$, we have
\beq
&& {\left( \frac{\chi_k+\nu k^2}{|\OB|k}\right) }^2 - \frac{(\chi_k+\nu
k^2)^2\sqrt{\chi_k^2+(\la+\nu)k^2\chi_k+\la\nu k^4}}{(|\OB|k)^3}
\times  \nn\\ && \arctan{\left(\frac{|\OB|k}{\sqrt{\chi_k^2+(\la+\nu)k^2\chi_k+\la\nu
k^4}}\right)} \longrightarrow {\left( \frac{\chi_k}{|\OB|k}\right) }^2.
\eeq
With these approximations, we have that under conditions (a), (b) and
(c),
\be
\alpha \approx -\frac{1}{2{\overline B}^2}\int^{+\infty}_0 dk [\tau_{v,k} 8\pi k^2C(k)  -
\tau_{b,k} 8\pi k^2H(k) ],
\ee
where $C(k)$ and $H(k)$ are related to kinetic helicity and current
helicity through (45) and (46). Note that this result is independent of
$\nu(=\la)$. (56) shows that $\alpha$ falls rapidly
as ${\overline B}$ increases. To further relate $\alpha$ with the
statistical properties of MHD turbulence, we introduce a vector $\bf
u^{(0)}$ that is the vector potential of velocity field $\v^{(0)}$
through $\nabla \times {\bf u}^{(0)} = \v^{(0)}$. Let ${\bf a}^{(0)}$
be the vector potential for $\b^{(0)}$. It is easy to find the relation
\be
\int 8\pi k^2C(k) dk = < {\bf u}^{(0)} \cdot \v^{(0)} >,
\ee
and
\be
\int 8\pi k^2H(k) dk = < \a^{(0)} \cdot \b^{(0)} >.
\ee
The right hand side of equation (58) is the magnetic helicity of the
turbulent field $\b^{(0)}$. With (57) and (58), we may
re-write (56) as
\be
\alpha = - \left(\frac{<{\bf u}^{(0)} \cdot \v^{(0)}>}{2{\overline
B}^2\tau_v} - \frac{<{\bf a}^{(0)} \cdot \b^{(0)}>}{2{\overline
B}^2\tau_b} \right).
\ee
(59) relates $\alpha$-effect with the dynamics of magnetic helicity of
turbulence. Because magnetic helicity is a conserved quantity in ideal
MHD, $\alpha$ coefficient derived above will be constrained by the
dynamics of magnetic helicity.

Finally, we consider how non-unit Prandtl number affects $\alpha$
dynamo. To aid our
discussion, we introduce the following quantities:
\be
R_{m,\omega}(k)=\frac{\omega_{k}}{\la {k}^2},
\ee
\be
R_{e,\omega}(k)=\frac{\omega_{k}}{\nu {k}^2},
\ee
\be
R_{m,\chi}(k)=\frac{\chi_{k}}{\la {k}^2},
\ee
\be
R_{e,\chi}(k)=\frac{\chi_{k}}{\nu {k}^2},
\ee
\be
\beta_1(k)=\frac{|\OB|k}{\omega_{k}},
\ee
\be
\beta_2(k)=\frac{|\OB|k}{\chi_{k}}.
\ee
These parameters are determined by the statistical properties at
different scales of the MHD turbulence of $\v^{(0)}$ and
$\b^{(0)}$. They are all functions of $k$. Each of the first four
quantities denotes the ratio of the kinetic or magnetic dissipation
time to the de-correlation time of velocity field or magnetic field,
therefore is similar to the definition of Reynolds numbers, which are
the ratios of nonlinear term to the dissipation term in the momentum
equation or the induction equation. $\beta_1$ and $\beta_2$ measure the
strength of large-scale magnetic field in the units of $v(k)$. Let
$t_v$ be the eddy turn-over time of the turbulence at the forcing scale
$l_0$. For these parameters at scale $l_0 = 1/k_0$, with the
experimental result (15), we have their relations to the conventional
Reynolds numbers as
\be
R_{m,\omega}(k_0)= R_m \frac{t_v}{\tau_v} \approx (3 \sim 5)R_m ,
\ee
\be
R_{e,\omega}(k_0)= R_e \frac{t_v}{\tau_v} \approx (3 \sim 5)R_e,
\ee
\be
R_{m,\chi}(k_0)= R_m \frac{t_v}{\tau_b} \approx (3 \sim 5)R_m,
\ee
\be
R_{e,\chi}(k_0)= R_e \frac{t_v}{\tau_b} \approx (3 \sim 5)R_e,
\ee
\be
\beta_1(k_0)= \frac{|\OB|}{v_0}\frac{\tau_v}{t_v} \approx (0.2
\sim 0.3)\frac{|\OB|}{v_0},
\ee
\be
\beta_2(k_0)= \frac{|\OB|}{v_0}\frac{\tau_b}{t_v} \approx (0.2
\sim 0.3)\frac{|\OB|}{v_0}.
\ee

Next, we express the general result of (44) using these statistical
quantities of the MHD turbulence. We first write
\be
\alpha = \int_0^{\infty} \alpha_k dk,
\ee
where
\be
\alpha_k = -\frac{1}{2}\left[\frac{2}{1+Pr}\Delta_v(k)\Pi_v(k) -
\frac{2Pr}{Pr+1}\Delta_b(k)\Pi_b(k) \right].
\ee
Here $Pr=\nu/\la$ is the magnetic Prandtl number. Define
$Z_M(k)=\frac{2}{1+Pr}\Pi_v(k)$, $Z_N(k)=\frac{2Pr}{Pr+1}\Pi_b(k)$; then
\be
\alpha_k=\frac{1}{2}\left[\Delta_v(k)Z_M - \Delta_b(k)Z_N\right].
\ee
With those quantities introduced in (60)-(65) and the definitions of
$Z_M$ and $Z_N$, we have
\beq
Z_M(k) &=& \frac{2}{1+Pr} \left\{ \left[ \left(\frac{1}{\beta_1
R_{e,\omega}}\right)^2 -\frac{\sqrt{1/Pr}}{(\beta_1 R_{e,\omega})^3}
\arctan{\frac{\beta_1 R_{e,\omega}}{\sqrt{1/Pr}}}\right] \r. \nn\\
&+& \l. \frac{R_{e,\omega}-1}{R_{e,\omega}+1}\left
[ {\left(\frac{R_{e,\omega}+1}{\beta_1 R_{e,\omega}}\right)}^2 -
\frac{(R_{e,\omega}+1)^2
\sqrt{R_{e,\omega}^2+(1/Pr+1)R_{e,\omega}+1/Pr}}{(\beta_1
R_{e,\omega})^3} \times \r. \r. \nn\\
&& \l. \l. \arctan{\left(\frac{\beta_1
R_{e,\omega}}{\sqrt{R_{e,\omega}^2+(1/Pr+1)R_{e,\omega}+1/Pr}}\right)}
\right] \right\},
\eeq
\beq
Z_N(k) &=& \frac{2Pr}{Pr+1} \left\{ \left[ \left(\frac{1}{\beta_2
R_{m,\chi}}\right)^2 -\frac{\sqrt{Pr}}{(\beta_2 R_{m,\chi})^3}
\arctan{\frac{\beta_2 R_{m,\chi}}{\sqrt{Pr}}}\right] \r. \nn\\
&+& \l. \frac{R_{m,\chi}-1}{R_{m,\chi}+1}\left
[ {\left(\frac{R_{m,\chi}+1}{\beta_2 R_{m,\chi}}\right)}^2 -
\frac{(R_{m,\chi}+1)^2
\sqrt{R_{m,\chi}^2+(Pr+1)R_{m,\chi}+Pr}}{(\beta_2
R_{m,\chi})^3} \times \r. \r. \nn\\
&& \l. \l. \arctan{\left(\frac{\beta_2
R_{m,\chi}}{\sqrt{R_{m,\chi}^2+(Pr+1)R_{m,\chi}+Pr}}\right)}
\right] \right\}.
\eeq
If we compare (74)
with the classical result given by (51), we find that the non-unit
Prandtl number will modify the $\alpha$ coefficient through the
complicated terms $Z_M$ and $Z_N$. For $Pr \rightarrow \infty$, we
expect that $Z_M \rightarrow 0$; therefore, the contribution to $\alpha$
from the velocity field will be small. But large $Pr$ may not switch off
the contribution of magnetic field to $\alpha$ because $Z_N$ can remain
moderate for $Pr \rightarrow \infty$. On the contrary, for $Pr
\rightarrow 0$, $Z_N \rightarrow 0$, so the main contribution to
$\alpha$ comes from $\Delta_v$. To further illustrate this, we plot
$Z_M$ and $Z_N$ at different $\beta_1$ and $\beta_2$ in Figure 1. $Z_M$ is
bounded above by $\frac{2}{3}$ and decreases as $1/\beta_1^2$. When
$R_{M,\omega}$ and $\beta_1$ are fixed, $Z_M$ achieves its maximum when
$Pr$ falls in a certain interval. Outside the interval, $Z_M$ falls
sharply to $0$. $Z_N$ has similar properties (in fact, $Z_N$ and $Z_M$ are
symmetric with respect to $Pr=1$). In Figure 2 we show other views of
$Z_M$ and $Z_N$ as functions of $R_{e,\omega}$ and $R_{m,\chi}$ when
$\beta_{1,2}(\equiv 0.0001)$ and $Pr$ are fixed. The curve for $Z_M$
with $Pr=1$ is already shown in Figure 1 of Field \etal(1999); however, as $Pr$
increases or decreases from 1, the distribution of $Z_M$ as a function
of $R_{e,\omega}$ changes dramatically: for small $Pr$ like $10^{-3}$,
$Z_M$ stays at its maximum value $\frac{2}{3}$ for all $R_{e,\omega}
\gtrsim 10^5$, compared to $Z_M \simeq 0$ for all $R_{e,\omega}
\gtrsim 10^5$ when $Pr = 10^3$, and $Z_M \simeq \frac{1}{3}$ for all
$R_{e,\omega} \gtrsim 10^5$ when $Pr=1$. But we must consider the
contribution from $Z_N$, too. In panel (b) of Figure 2 we plot a few
curves of $Z_N$ as a function of $R_{m,\chi}$. Close investigation of
panels (a) and (b) in Figure 2 shows that $Z_M$ and $Z_N$ may
complement each other such that if the helicity terms $|\Delta_v| \sim
|\Delta_b|$(see equation (74)), $\alpha$ can maintain a non-trivial
value for a large range of $Pr$, and this is shown in Figure 3, where
we plot the regimes where $Z_M \geq 0.1$ or $Z_N \geq 0.1$ for
$\beta_1 = \beta_2 = 0.0001$. The vertically shaded area by solid lines
is the regime of $R_{e,\omega}$ and $R_{m,\omega}$ where the
contribution to $\alpha$ from velocity field, $Z_M$, is no less than
0.1 for $\beta_1=0.0001$, while the horizontally shaded area by
dash-dot lines is the regime of $R_{e,\chi}$ and $R_{m,\chi}$ where the
contribution to $\alpha$ from magnetic field, $Z_N$, is no less than
0.1 for $\beta_2=0.0001$. The case with stronger large-scale magnetic
field is shown in Figure 4 with $\beta_1 = \beta_2 = 1$. An
$(R_{e,\omega}, R_{m,\omega}, R_{e,\chi}, R_{m,\chi})$ combination
corresponds to two points on the plane of Figure 3 or Figure 4, one for
$(R_{e,\omega}, R_{m,\omega})$, the other for $(R_{e,\chi},
R_{m,\chi})$. If either of these two points falls within its
corresponding shaded area, or both do, we may obtain a non-vanishing
$\alpha$ which by (74) satisfies 
\be
|\alpha| \gtrsim
\min\{0.05|\Delta_v|,0.05|\Delta_b|,0.05||\Delta_v|-|\Delta_b||\}
\ee
at $\beta_1 = \beta_2 = 0.0001$.

\section{Discussion}
In our derivation for $\alpha$, the de-correlation time for
velocity field at scale $1/k$, $1/\omega_{v,k}$, can be different
from the de-correlation time of magnetic field, $1/\chi_{b,k}$, at the
same scale. These two parameters are the statistical properties of the MHD
turbulence, and can be measured in numerical simulations or
experiments. Under different circumstances, these two parameters can be
different. They are affected by molecular diffusion and magnetic
diffusion, i,e., $\nu$ and $\la$, by the properties of external forcing
term $\f$, and by the nonlinear interactions between different modes of
velocity field and magnetic field at different scales.

We also generalize the derivation of the dynamo $\alpha$-coefficient
to include non-unit Prandtl number. Prandtl number is an important physical
parameter. For many astrophysics systems, $Pr \neq 1$. For example, in
the solar convection zone, $Pr \sim 10^{-6} - 10^{-2}$ (Childress and
Gilbert, 1995); while in the partially ionized warm gas of the
interstellar medium $Pr \sim 10^{15}$ (Kinney \etal, 2000). Many past
and recent discussions on $\alpha$ dynamo assume unit Prandtl number,
which may not appropriate for real astrophysics systems. Our analysis
of non-unit Prandtl number case shows that, for astrophysical systems
such as solar convection zone  where $Pr$ can be very small, the
contribution to $\alpha$ dynamo from kinetic helicity can be much
stronger than the contribution from current helicity. In this limit,
our result can be reduced to the classical result of kinematic
mean-field theory, i.e., the current helicity term does not play
important role in determining $\alpha$. This can be seen from (73) by
setting $Pr \rightarrow 0$. However, for the opposite limit, i.e., $Pr
\rightarrow \infty$, the contribution to $\alpha$ from current helicity
can be much more important than that from kinetic helicity. This means
for astrophysical systems such as warm interstellar medium, dynamo
$\alpha$ effect will strongly depend on the current helicity of
magnetic field of the system instead of the kinetic helicity.

Another possible feature of $\alpha$ dynamo considered in this work is
that, regardless of large or small Prandtl number, there is no simple
relation as
\be
\alpha \propto \frac{\alpha_0}{1+R_m {\overline B}^2/v_0^2}
\ee
for $R_m \gg 1$. Rather, nonlinear interactions of the turbulence
velocity field $\v^{(0)}$ and $\b^{(0)}$ will lead to finite
de-correlation times of $\v^{\prime}$ and $\b^{\prime}$ so that even for
$R_m, R_e \gg 1$, the electromotive force is not significantly
reduced. This can be easily seen from the plot in Figures 3 and
4. $\alpha$ calculated in the shaded areas (horizontal or vertical) is
not dramatically quenched. Such shaded areas span large parameter space
of $(R_{e,\omega}, R_{m,\omega}, R_{e,\chi}, R_{m,\chi})$. Hence, we
believe that for large kinetic and magnetic Reynolds numbers, the
following relation is approximately true 
\be
\alpha \propto \frac{\alpha_0}{{\overline B}^2/v_0^2}
\ee
for moderate or large ${\overline B}$ (see expressions (56) or
(59)). Here, for $Pr \sim 1$, $\alpha_0 = -\frac{\tau_v}{3}<\v \cdot
\nabla \times \v> + \frac{\tau_b}{3}<\b \cdot \nabla \times \b>$; for
$Pr \ll 1$, $\alpha_0 = -\frac{\tau_v}{3}<\v \cdot \nabla \times \v>$;
and for $Pr \gg 1$, $\alpha_0 = \frac{\tau_b}{3}<\b \cdot \nabla \times
\b>$. Relation (79) is close to relation (9), which Kraichnan proposed in
1979. Indeed, our model is closely related to, but not exactly the
same as, Kraichnan's oversimplified model. A detailed comparison between
our model and the model by Kraichnan has been given in appendix B.

The derivation of dynamo $\alpha$ effect in this work should be
considered as an extension of the kinematic MFE to the dynamic
regime. We use two-scale approach, but we treat the velocity field and
the magnetic field on equal footing. Our assumption of short time
de-correlation (see relation (14)) has been proved by both experiments and
numerical simulations. There are, however,  the circumstances that
condition (14) is not valid, such as the small-scale magnetic field
amplification problem considered by Kulsrud and Anderson (1992). In
their work, when magnetic field is weak so back reaction can be
ignored, the de-correlation time of magnetic field can be long compared with
the line stretching time; therefore, the nonlinear terms must be
kept. Kulsrud and Anderson (1992) did exactly that. They assumed that
the smallest time in the frame of their work is the turn-over time of
the smallest eddies, $\tau_{max}$, and the dominant effect is the line
stretching by turbulent eddies. But their results can be valid only
for small back reaction. In our model, back reaction can be strong, and the
smallest time is the de-correlation time of the $\OB$-dependent
$\v^{\prime}$ and $\b^{\prime}$.

It is also possible to incorporate into our model the nonlinear effects
from those nonlinear terms that we omitted in section 2, if the
de-correlation times of velocity and magnetic fields are long compared
to eddy turn-over time. To aid our discussion, we focus on the case
that $\OB$ is strong. In this case, the MHD turbulence can be treated
as a bath of Alfv\'{e}n waves. In appendix B, we argue that in the
limit of strong large-scale magnetic field $\OB$, $\v^{\prime}$ and
$\b^{\prime}$ can be regarded as $\{ \v^{(0)}, \b^{(0)} \}$-driven,
Alfv\'{e}nic wave-like motions. Because $\v^{(0)}$ and $\b^{(0)}$ can
be de-correlated within $\tau_{dcor} < \tau_{eddy}$, the $\{ \v^{(0)},
\b^{(0)} \}$-driven $\v^{\prime}$ and $\b^{\prime}$ will also be
de-correlated within $\tau_{dcor} < \tau_{eddy}$. Our understanding of
MHD turbulence is as follows: for statistically steady state of the MHD
turbulence, the {\it net} effect of the nonlinear interactions between
different modes of such wave-like motions, is that the kinetic energy
and the magnetic energy will cascade from forcing scale down to
dissipation scales and be converted to heat. Therefore, to incorporate
the nonlinear interactions between different modes of Alfv\'{e}nic
wave-like quantities $\v^{\prime}$ and $\b^{\prime}$ into our model, we
approximate the nonlinear terms with linear, effective turbulent mixing
terms, and re-write equation (17) and (18) as
\be
\left( -i\omega + \nu k^2 + \sigma_{nd}k^2  \right)
{\hat{v}}^{\prime}_j  + \gamma_{nd}k^2{\hat{b}}^{\prime}_j = i(\k \cdot
\OB)({\hat{b}}^{\prime}_j + {\hat{b}}^{(0)}_j),
\ee
\be
\left( -i\omega + \la k^2 + \eta_{nd}k^2 \right) {\hat{b}}^{\prime}_j +
\varpi_{nd}k^2{\hat{v}}^{\prime}_j = i(\k \cdot
\OB)({\hat{v}}^{\prime}_j + {\hat{v}}^{(0)}_j).
\ee
Here we approximate the nonlinear terms in the (Fourier transformed)
equations for ${\hat \v}^{\prime}$ and ${\hat \b}^{\prime}$ with
$\sigma_{nd}k^2 {\hat \v^{\prime}} + \gamma_{nd}k^2 {\hat \b^{\prime}}$
and $\eta_{nd}k^2 {\hat \b^{\prime}} + \varpi_{nd}k^2 {\hat
\v^{\prime}}$ (see Appendix B for more discussions). All of the
nonlinear damping factors, $\sigma_{nd}$, $\gamma_{nd}$, $\eta_{nd}$ and
$\varpi_{nd}$, are functions of $k$. These damping factors can be
calculated with closure theory (Biskamp, 1994; see also Chen and
Montgomery, 1987, or Kraichnan, 1979). In the limit of large $\overline
B$, compared with terms involving $i\k \cdot \OB$, the nonlinear
damping factors $\gamma_{nd}$ in (80) and $\varpi_{nd}$ in (81) have
significant contributions to the damping of turbulence only near the
plane that is perpendicular to $\OB$; therefore, their contributions to
the integrations with respect to $\theta = \angle(\k, \OB)$ are small
and we omit them to get the first order approximation. Under such
considerations, molecular viscosity $\nu$ and magnetic diffusivity
$\la$ in our calculations for $\alpha$ in section 4 must be replaced by
$\nu + \sigma_{nd}$ and $\la + \eta_{nd}$, respectively, and the
magnetic Prandtl number in expression (73) has to be replaced by
\be
Pr_{nd}(k) = \frac{\nu + \sigma_{nd}(k)}{\la + \eta_{nd}(k)},
\ee
and the $\alpha_k$ in (73) must be re-written as
\be
\alpha_{k,nd} =
-\frac{1}{2}\left[\frac{2}{1+Pr_{nd}(k)}\Delta_v(k)\Pi_{v,nd}(k) -
\frac{2Pr_{nd}(k)}{Pr_{nd}(k)+1}\Delta_b(k)\Pi_{b,nd}(k) \right]. 
\ee
Here $\Pi_{v,nd}(k)$ and $\Pi_{b,nd}(k)$ can be obtained by replacing
$\nu$ by $\nu + \sigma_{nd}(k)$ and by replacing $\la$ by $\la +
\eta_{nd}(k)$ in expressions (42) and (43). The approximation to
nonlinear effects with $\sigma_{nd}(k)$ and $\eta_{nd}(k)$ are simple,
straightforward, but crude. Detailed analysis of the exact forms of
$\sigma_{nd}(k)$ and $\eta_{nd}(k)$ with closure theory or other
analytic methods can be complicated and is beyond the scope of this
work.

It has been realized that the conservation of magnetic helicity is
related to the $\alpha$ dynamo process. Many authors in the literature
have suggested that for large magnetic Reynolds number, $\alpha$ effect
is suppressed. Our model does not consider the conservation of magnetic
helicity. This is because we consider the $\alpha$ effect as a result
of the nonlinear interaction between homogeneous, isotropic MHD
turbulence driven by an external force and a large-scale magnetic field,
$\OB$. In our model, the external forcing term has the freedom to drive
turbulence that will interact with $\OB$ to produce a non-zero $\alpha$
dynamo in large $R_m$ limit, while in the work by Gruzinov and Diamond
(1994, 1995, 1996; see also Zeldovich, 1957), the velocity field is
simply assumed to be given and no external energy sources are
considered. This is not true for astrophysical systems where
``...kinetic energy is constantly being injected hydrodynamically [at
the forcing scales].''(Kulsrud, 1999). Note that the result that
$\alpha$ effect should be quenched due to the conservation of magnetic
helicity in the large $R_m$ limit is obtained by assuming a steady
state and a closed system. As Blackman and Field(2000) point out, the
assumption that a system is closed forbids a non-zero net flux of
magnetic helicity to flow into or out of the system. In real
astrophysical systems like the Sun, the boundary is open, and magnetic
helicity can flow through it. Moreover, the assumption of steady state
is not valid in many astrophysical systems where transient, impulsive
activities, such as solar flares, can happen(Chou and Field,
2000). With no assumptions of closeness or stationarity of the system,
we believe the $\alpha$ dynamo can operate under the nonlinear
interaction between $\OB$ and the MHD turbulence, thus can be
determined by $\OB$ and the statistical properties of the turbulence. A
more thorough, though complicated, model of $\alpha$ dynamo than ours
should incorporate the boundary effects into the discussions.  

\section{Conclusion}
We generalize the derivation of the dynamo $\alpha$-coefficient to include
non-unit Prandtl number and different de-correlation times of
velocity/magnetic field. Our formula gives $\alpha$ as a functional of
the statistical properties of ${\b}^{(0)}$ and ${\v}^{(0)}$, and it is
also a function of several parameters introduced in section 4.  We
confirm the fundamental results of Field, Blackman and Chou(1999), that
$\alpha$ does not vanish at very high Reynolds numbers, but rather
obtains a finite value. There are two parts in the expression of
$\alpha$: one is from velocity field, the other is from magnetic
field. Non-trivial contribution of either of these two parts can be
achieved only when the Prandtl number $Pr$ falls in certain regimes,
but the total contribution to $\alpha$ can be insensitive to $Pr$ if
kinetic helicity and current helicity are comparable.    

\acknowledgements
I am grateful to Prof. George Field for many stimulating discussions. I
also benefit from my discussions with E.G. Blackman, A. Brandenberg,
B. Chandran, S. Cowley, R. Kulsrud, D. Balsra. 

\appendix
\section{Calculation of Electromotive Force}
We start from the calculation of correlation function of velocity field
and magnetic field. If we plug (19) and (20) into (25) in section 3, we have
\beq
\lra{v_l(\x,t)b_j(\x,t)} &=& \int \! \int d{\k}^{\prime}
d{\omega}^{\prime} {\left[\int \! \int d\k d\omega e^{i(\k \cdot \x
- \omega t)}
\lra{{\hat{v}}_l({\k}^{\prime},{\omega}^{\prime}){\hat{b}}_j(\k-{\k}^{\prime}
,\omega-{\omega}^{\prime})}\right]} \nn\\
&=& \int \! \int d{\k}^{\prime} d{\omega}^{\prime}
\{[1-\zeta(-{\k}^{\prime},-{\omega}^{\prime})-\zeta({\k}^{\prime},
{\omega}^{\prime}) + \nn\\
&& \zeta({\k}^{\prime},{\omega}^{\prime})
\zeta(-{\k}^{\prime},-{\omega}^{\prime})]{\hat{P}}^{00}_{lj}(-{\k}^
{\prime},-{\omega}^{\prime}) + \nn\\
&&
[\xi_{\la}(-{\k}^{\prime},-{\omega}^{\prime})-\zeta({\k}^{\prime},{\omega}^
{\prime}){\xi}_{\la}(-{\k}^{\prime},-{\omega}^{\prime})]{\hat{Q}}^{00}_{lj}(-
{\k}^{\prime},-{\omega}^{\prime})] + \nn\\
&&
[\xi_{\nu}({\k}^{\prime},{\omega}^{\prime})-\zeta(-{\k}^{\prime},-{\omega}^
{\prime}){\zeta}_{\nu}({\k}^{\prime},{\omega}^{\prime})]{\hat{S}}^{00}_{lj}(-
{\k}^{\prime},-{\omega}^{\prime})] + \nn\\
&&
[\xi_{\nu}({\k}^{\prime},{\omega}^{\prime})\xi_{\la}(-{\k}^{\prime},-{\omega}^{\prime})]{\hat{P}}^{00}_{jl}({\k}^{\prime},{\omega}^{\prime})\}. 
\eeq
Note that
$[1-\zeta(-\k,-\omega)-\zeta(\k,\omega)+\zeta(\k,\omega)\zeta(\-k,-\omega)]$
and $[\xi_{\nu}(\k,\omega)\xi_{\la}(-\k,-\omega)]$ are {\it even} functions
of $\k$, while
$[\xi_{\la}(-\k,-\omega)-\zeta(\k,\omega)\xi_{\la}(-\k,-\omega)]$ and
$[\xi_{\nu}(\k,\omega)-\zeta(-\k,-\omega)\xi_{\nu}(\k,\omega)]$ are
{\it odd}. Because of the homogeneity and isotropy of the turbulence,
${\hat{Q}}^{00}_{lj}(-\k,\omega)={\hat{Q}}^{00}_{jl}(\k,\omega)$;
therefore the antisymmetric tensor
${\hat{Q}}^{00}_{lj}(\k,\omega)-{\hat{Q}}^{00}_{jl}(\k,\omega)$ is an {\it
odd} function of $\k$, as are
${\hat{P}}^{00}_{lj}(\k,\omega)-{\hat{P}}^{00}_{jl}(\k,\omega)$ and
${\hat{S}}^{00}_{lj}(\k,\omega)-{\hat{S}}^{00}_{jl}(\k,\omega)$, so
many terms in (A1) vanish. By these, we have the electromotive force
$\lra{\v \times \b}_n = M_n + N_n$. $M_n$ and $N_n$ are given in equations (34)
and (35) of main text.

With the definitions of $\xi_{\la}, \xi_{\nu}$ and $\zeta$ in (21), (22) and
(23), we can rewrite $M_n$ and $N_n$ as
\be
M_n=\int \int d\k d\omega \left\{ \frac {\frac {i(\k \cdot
\OB)}{-i\omega+\la k^2}}{\left[1+\frac{(\k \cdot \OB)^2}{(-i\omega +
\la k^2)(-i\omega + \nu k^2)}\right] \left[1+\frac{(\k \cdot
\OB)^2}{(i\omega + \la k^2)(i\omega + \nu k^2)}\right]}
\right\}\ep_{nlj} {\hat{Q}}^{00}_{lj}(\k,\omega)
\ee
and
\be
N_n=\int \int d\k d\omega \left\{ \frac {\frac {-i(\k \cdot
\OB)}{i\omega+\nu k^2}}{\left[1+\frac{(\k \cdot \OB)^2}{(-i\omega +
\la k^2)(-i\omega + \nu k^2)}\right] \left[1+\frac{(\k \cdot
\OB)^2}{(i\omega + \la k^2)(i\omega + \nu k^2)}\right]}
\right\}\ep_{nlj} {\hat{S}}^{00}_{lj}(\k,\omega).
\ee
Define
\be
\omega_{1,2}=\frac{-i(\la+\nu)k^2 \pm \sqrt{-(\la-\nu)^2 k^4 + 4(\k
\cdot \OB)^2}}{2},
\ee
\be
\omega_{3,4}=\frac{i(\la+\nu)k^2 \pm \sqrt{-(\la-\nu)^2 k^4 + 4(\k
\cdot \OB)^2}}{2}.
\ee
In the following, we assume the square roots in (A4) and (A5) are
real (otherwise, see Chou and Fish, 1999). Under the homogeneity and
isotropy assumptions made in section 3, $M_n$ can be computed as
\be
M_n=\frac{\ep_{nlj}}{2\pi}\int d{\k} {\hat{X}}^{00}_{lj}(\k)\left[-(\k
\cdot \OB)\right] \int^{+\infty}_{-\infty} d\varphi T(\varphi)
\int^{+\infty}_{-\infty} \frac{e^{i\omega\varphi}(\omega-i\la
k^2)({\la}^2k^4 +
\omega^2)d\omega}{(\omega-\omega_1)(\omega-\omega_2)(\omega-\omega_3)
(\omega-\omega_4) }.
\ee
Contour integration of (A6) gives
\be
M_n=\frac{1}{2\pi}\int d{\k} \ep_{nlj} {\hat{X}}^{00}_{lj}(\k)[-(\k
\cdot \OB)](2 \pi i)[U-V]
\ee
where
\be
U=\int^{+\infty}_{0}d\varphi e^{-\omega_k \varphi}(e^{i\omega_3
\varphi} Y1 + e^{i\omega_4 \varphi} Y2),
\ee
\be
V=\int^{0}_{-\infty}d\varphi e^{\omega_k \varphi}(e^{i\omega_1
\varphi} W1 + e^{i\omega_2 \varphi} W2),
\ee
\be
Y1=\frac{(\omega_3-i\la k^2)({\omega}_3^2+\nu^2
k^4)}{\left[i(\la+\nu)k^2\right][2\omega_3]\left[\sqrt{-(\la-\nu)^2
k^4+4(\k \cdot \OB)^2}\right]},
\ee
\be
Y2=\frac{(\omega_4-i\la k^2)({\omega}_4^2+\nu^2
k^4)}{\left[i(\la+\nu)k^2\right][2\omega_4]\left[-\sqrt{-(\la-\nu)^2
k^4+4(\k \cdot \OB)^2}\right]},
\ee
\be
W1=\frac{(\omega_1-i\la k^2)({\omega}_1^2+\nu^2
k^4)}{\left[-i(\la+\nu)k^2\right][2\omega_1]\left[\sqrt{-(\la-\nu)^2
k^4+4(\k \cdot \OB)^2}\right]},
\ee
\be
W2=\frac{(\omega_2-i\la k^2)({\omega}_2^2+\nu^2
k^4)}{\left[-i(\la+\nu)k^2\right][2\omega_2]\left[-\sqrt{-(\la-\nu)^2
k^4+4(\k \cdot \OB)^2}\right]}.
\ee
Because $Y1,Y2,W1,W2$ are not functions of $\varphi$, $U$ and $V$ can be
calculated as
\beq
U &=& Y1\int^{+\infty}_0 d\varphi e^{-\omega_k \varphi} e^{i\omega_3
\varphi} + Y2\int^{+\infty}_0 d\varphi e^{-\omega_k \varphi} e^{i\omega_4
\varphi} \nn\\
&=& Y1 \frac{1}{\omega_k-i\omega_3} + Y2 \frac{1}{\omega_k-i\omega_4}
\nn\\
&=&
\frac{\sigma}{\left[2i(\la+\nu)k^2\right]\left[\sqrt{-(\la-\nu)^2k^4+4(\k
\cdot \OB)^2}\right]}
\eeq
where
\be
\sigma=\frac{(\omega_3-i\la k^2)(\omega^2_3+\nu^2
k^4)}{\omega_3(\omega_k-i\omega_3)}-\frac{(\omega_4-i\la
k^2)(\omega^2_4+\nu^2 k^4)}{\omega_4(\omega_k-i\omega_4)};
\ee
and
\beq
V &=& W1\int_{-\infty}^0 d\varphi e^{\omega_k \varphi} e^{i\omega_1
\varphi} + W2\int_{-\infty}^0 d\varphi e^{\omega_k \varphi} e^{i\omega_2
\varphi} \nn\\
&=& W1 \frac{1}{\omega_k+i\omega_1} + W2 \frac{1}{\omega_k+i\omega_2}
\nn\\
&=&
\frac{\gamma}{\left[-2i(\la+\nu)k^2\right]\left[\sqrt{-(\la-\nu)^2k^4+4(\k
\cdot \OB)^2}\right]}
\eeq
where
\be
\gamma=\frac{(\omega_1-i\la k^2)(\omega^2_1+\nu^2
k^4)}{\omega_1(\omega_k+i\omega_1)}-\frac{(\omega_2-i\la
k^2)(\omega^2_2+\nu^2 k^4)}{\omega_2(\omega_k+i\omega_2)}.
\ee
Because of the following five identities:
\be
\omega_1\omega_2 = \omega_3\omega_4 = -\la\nu k^4 -(\k \cdot \OB)^2,
\ee
\be
\omega_1 +\omega_2 = -i(\la+\nu)k^2=-(\omega_3+\omega_4),
\ee
\be
\omega_k+i\omega_1 = \omega_k - i\omega_3,
\ee
\be
\omega_k+i\omega_2 = \omega_k - i\omega_4,
\ee
\be
\omega_1 - \omega_2 = \omega_3 - \omega_4 = \sqrt{-(\la-\nu)^2 k^4 +
4(\k \cdot \OB)^2},
\ee
with (A4) and (A5), we have then
\beq
U-V &=& \left[ \frac{\la\nu^2 k^4}{(\la+\nu)\omega_k} -
\frac{\omega_k\la}{\la+\nu} \right]
\frac{1}{{\left(\sqrt{\omega^2_k+(\la+\nu)k^2\omega_k+\la\nu
k^4}\right)}^2+(k|\OB|\cos\theta)^2} \nn\\
&-& \frac{\la\nu^2 k^4}{(\la+\nu)\omega_k} \frac{1}{\la\nu
k^2+(k|\OB|\cos\theta)^2}.
\eeq
Here $\theta$ is the angle between vectors $\k$ and $\OB$. With (36)
and (37) in section 3 of main text, we have that
$\ep_{nlj}{\hat{X}}^{00}_{lj}(\k)=i\ep_{nlj}\ep_{ljk}k_kC(k)=
i2k_nC(k)$, and
$\ep_{nlj}{\hat{O}}^{00}_{lj}(\k)=i\ep_{nlj}\ep_{ljk}k_kH(k)=
i2k_nH(k)$. Let $\rho=\sqrt{\omega^2_k+(\la+\nu)k^2\omega_k+\la\nu
k^4}$, $y=k|\OB|\cos\theta$, and define $\Theta(y)$ as
\be
\Theta(y)=\frac{\la \nu^2 k^4-\la \omega_k^2}{(\la + \nu) \omega_k}
\frac{1}{\rho^2 + y^2} - \frac{\la \nu^2 k^4}{(\la + \nu)
\omega_k}\frac{1}{\la\nu k^2 + y^2}.
\ee
Then $M_n$ can be rewritten as
\be
M_n=-4\pi\int^{+\infty}_0 dk k^4 C(k) {\overline B}_n \int^{\pi}_0 \cos^2\theta \, d(\cos\theta) \, \Theta(k|\OB|\cos\theta).
\ee
Note that
\beq \int_0^{\pi}\cos^2\theta d(\cos\theta) \Theta
(k|\OB|\cos\theta) &=& \frac{\la \omega^2_k - \la\nu^2
k^4}{(\la+\nu)\omega_k}\frac{1}{(k|\OB|)^3} \int^{k|\OB|}_{-k|\OB|}
\frac{y^2}{\rho^2+y^2} dy \nn\\ &+& \frac{\la\nu^2 k^4}{(\la+\nu)\omega_k}
\frac{1}{(k|\OB|)^3} \int^{k|\OB|}_{-k|\OB|} \frac{y^2}{(\sqrt{\la\nu
k^4})^2+y^2} dy \nn\\
&\equiv& \frac{1}{\omega_k}\frac{2\la}{\la+\nu}\Pi_v(k)
\eeq
where
\beq
\Pi_v(k) &=& \left[{\left(\frac{\nu k^2}{|\OB|k}\right)}^2 - \frac{\nu^2
k^4 \sqrt{\la\nu
k^4}}{(|\OB|k)^3}\arctan\left(\frac{|\OB|k}{\sqrt{\la\nu k^4}}\right)
\right] \nn\\ &+& \frac{\omega_k-\nu k^2}{\omega_k+\nu k^2} \left
[ {\left( \frac{\omega_k+\nu k^2}{|\OB|k}\right) }^2 - \frac{(\omega_k+\nu
k^2)^2\sqrt{\omega_k^2+(\la+\nu)k^2\omega_k+\la\nu
k^4}}{(|\OB|k)^3} \times \r. \nn\\ && \l. \arctan{\left(\frac{|\OB|k}{\sqrt{\omega_k^2+(\la+\nu)k^2\omega_k+\la\nu
k^4}}\right)}\right].
\eeq
Let
\be
\Delta_v(k)\equiv \frac{8\pi k^4C(k)}{\omega_k},
\ee
\be
\Delta_b(k)\equiv \frac{8\pi k^4H(k)}{\chi_k};
\ee
therefore we have
\be
M_n=-\frac{1}{2} \int^{+\infty}_0 dk \left
[ \frac{2\la}{\la+\nu}\Delta_v(k)\Pi_v(k)\right]{\ob}_n.
\ee
In parallel, we have
\be
N_n=\frac{1}{2} \int^{+\infty}_0 dk \left
[ \frac{2\nu}{\nu+\la}\Delta_b(k)\Pi_b(k)\right]{\ob}_n,
\ee
where
\beq
\Pi_b(k) &=& \left[{\left(\frac{\la k^2}{|\OB|k}\right)}^2 - \frac{\la^2
k^4 \sqrt{\nu\la
k^4}}{(|\OB|k)^3}\arctan\left(\frac{|\OB|k}{\sqrt{\nu\la k^4}}\right)
\right] \nn\\ &+& \frac{\chi_k-\la k^2}{\chi_k+\la k^2} \left
[ {\left( \frac{\chi_k+\la k^2}{|\OB|k}\right) }^2 - \frac{(\chi_k+\la
k^2)^2\sqrt{\chi_k^2+(\nu+\la)k^2\chi_k+\nu\la
k^4}}{(|\OB|k)^3} \times \r. \nn\\ && \l. \arctan{\left(\frac{|\OB|k}{\sqrt{\chi_k^2+(\nu+\la)k^2\chi_k+\nu\la
k^4}}\right)}\right].
\eeq
We finally achieve an expression for the electromotive force 
\be
<\v \times \b> = -\frac{1}{2} \int_0^{+\infty}dk
\left[\frac{2\la}{\la+\nu}\Delta_v(k)\Pi_v(k) -
\frac{2\nu}{\nu+\la}\Delta_b(k)\Pi_b(k) \right] \OB.
\ee

\section{Alfv\'{e}n wave-like properties of $\{\v^{\prime}, \b^{\prime} \}$
in the presence of strong $\OB$}
In the presence of a strong large-scale magnetic field $\OB$, MHD
turbulence takes the form of random Alfv\'{e}n waves propagating along
$\OB$ (Kraichnan 1965). This can be seen from (12) and (13). Indeed, if
we re-write (12) and (13) as
\be
\part_t {\v}^{\prime} = \OB \cdot \nabla {\b}^{\prime} + \{ \nu {\nabla}^2
{\v}^{\prime} + \OB \cdot \nabla {\b}^{(0)} \},
\ee
\be
\part_t {\b}^{\prime} = \OB \cdot \nabla {\v}^{\prime} + \{ \lambda
{\nabla}^2 {\b}^{\prime} + \OB \cdot \nabla {\v}^{(0)} \},  
\ee
we find that without the terms in brackets, equations (B1) and (B2)
have solutions of linear Alfv\'{e}n waves. These Alfv\'{e}n waves are damped
through the dissipation terms. External energy sources, represented by the
forcing term $\f$ in the momentum equation (11), will drive a MHD
turbulence represented by $\{\v^{(0)}, \b^{(0)}\}$, which in turn stretch
the large-scale magnetic field $\OB$ to drive the Alfven-wave-like
patterns of $\{\v^{\prime}, \b^{\prime}\}$. Such physical processes and
the damping due to $\{\nu, \la\}$ are represented by the terms in bracket in
equations (B1) and (B2). Therefore, in the presence of strong $\OB$,
$\{\v^{\prime}, \b^{\prime}\}$ are Alfv\'{e}nic wave-like motions that are
driven by $\{\v^{(0)}, \b^{(0)}\}$ and damped by $\{\nu, \la\}$
terms. To the degree that the condition (14) is not exactly obeyed,
such $\{\v^{(0)}, \b^{(0)}\}$-driven, $\{\nu, \la\}$-damped,
Alfv\'{e}nic wave-like motions can be further damped by nonlinear wave-wave
interactions, as discussed by Kraichnan (1979). In the limit of large
$\overline B$, if we use linear terms $\sigma_{nd}k^2$,
$\gamma_{nd}k^2$, $\eta_{nd}k^2$ and $\varpi_{nd}k^2$ to approximate
nonlinear damping terms\footnote{There terms are, using our notations
in section 2 of main text, the following: ${\b}^{\prime}
\cdot \nabla {\b}^{\prime}$, ${\b}^{\prime} \cdot \nabla {\b}^{(0)}$,
${\b}^{\prime} \cdot \nabla {\v}^{\prime}$, ${\b}^{\prime} \cdot
\nabla {\v}^{(0)}$, ${\v}^{\prime} \cdot \nabla {\b}^{\prime}$,
${\v}^{\prime} \cdot \nabla {\b}^{(0)}$, ${\v}^{\prime} \cdot \nabla
{\v}^{\prime}$, ${\v}^{\prime} \cdot \nabla {\v}^{(0)}$, ${\b}^{(0)}
\cdot \nabla {\b}^{\prime}$, ${\b}^{(0)} \cdot \nabla {\v}^{\prime}$,
${\v}^{(0)} \cdot \nabla {\v}^{\prime}$, and ${\v}^{(0)} \cdot \nabla
{\b}^{\prime}$.}, we may write, in Fourier space, the following
two equations for $\v^{\prime}$ and $\b^{\prime}$ 
\be
\left( -i\omega + \nu k^2 + \sigma_{nd}k^2  \right)
{\hat{v}}^{\prime}_j + \gamma_{nd}k^2{\hat{b}}^{\prime}_j = i(\k \cdot
\OB)({\hat{b}}^{\prime}_j + {\hat{b}}^{(0)}_j),
\ee
\be
\left( -i\omega + \la k^2 + \eta_{nd}k^2 \right) {\hat{b}}^{\prime}_j +
\varpi_{nd}k^2{\hat{v}}^{\prime}_j = i(\k \cdot
\OB)({\hat{v}}^{\prime}_j + {\hat{v}}^{(0)}_j).
\ee
These two equations are modifications to equations (B1) and
(B2). Following Kraichnan (1979), we consider single mode solutions of
(B3) and (B4):
\be
{\hat \b}^{\prime}(\k) \propto \left( \k \cdot \OB \right) e^{-iV_Akt-ir_b}
e^{-\Gamma_b(k) k^2t } \left(e^{i\phi_{b,k}t} \b^{(0)}(\k) -
e^{i\phi_{v,k}t} \v^{(0)}(\k) \right),
\ee
\be
{\hat \v}^{\prime}(\k) \propto \pm \left( \k \cdot \OB \right) e^{\mp
iV_Akt \mp ir_v} e^{-\Gamma_v(k) k^2t}
\left(e^{i\phi^{\prime}_{b,k}t} \b^{(0)}(\k) -
e^{i\phi^{\prime}_{v,k}t} \v^{(0)}(\k) \right).
\ee
Here $V_A = {\overline B}$ is the Alfv\'{e}n speed in the unit of $(4\pi
\rho)^{1/2}$. $r_b$ and $r_v$ are phases determined by initial
conditions. $\Gamma_{b(v)}(k)$ are linear combinations of the damping
effects $\la, \nu, \sigma_{nd}, \eta_{nd}, \gamma_{nd}$ and
$\varpi_{nd}$, i.e., damping effects from both nonlinear mode-mode
interactions and physical viscosity/diffusivity. $\phi^{(\prime)}_{v,k}
= g^{(\prime)}_{v,k} + \mbox{i}\omega_{v,k}$ is the phase of
$\v^{(0)}$, while $\phi^{(\prime)}_{b,k} = g^{(\prime)}_{b,k} +
\mbox{i}\chi_{b,k}$ is the phase of $\b^{(0)}$. Note that depending on
the initial condition, $g_{v(b),k}$ may not be equal to
$g^{\prime}_{v(b),k}$. Relations (B5) and (B6) tell us the following
points: (1) $\v^{\prime}$ and $\b^{\prime}$ are Alfv\'{e}n waves in
character; yet, (2) these Alfv\'{e}n waves are driven by $\{\v^{(0)},
\b^{(0)}\}$, which are in turn driven by external forcing term $\f$; (3) these
$\{\v^{(0)}, \b^{(0)}\}$-driven Alfv\'{e}n waves are damped in three
ways: (A) molecular viscosity $\nu$ and magnetic diffusivity $\la$, and (B)
nonlinear interactions between different modes of $\{\v^{\prime}$,
$\b^{\prime}\}$ approximated by parameters $\sigma_{nd}, \eta_{nd},
\gamma_{nd}$ and $\varpi_{nd}$, and (C) the de-correlation of
$\{\v^{(0)}, \b^{(0)}\}$ approximated by $\omega_{v,k}$ and
$\chi_{b,k}$. The $\alpha$ dynamo calculated from (B3) and (B4) in the
limit of large $\overline B$ is
\be
\alpha = \int_0^{\infty} \alpha_{k,nd} dk
\ee
where $\alpha_{k,nd}$ is given by (83).

Kraichnan (1979) demonstrated the $\alpha$ effect in the strong-field
limit where the turbulence consists of Alfv\'{e}n waves propagating along
$\OB$. He considered the following linearized single mode helical
standing wave,
\be
\v = u_0 \cos(V_A kt) \left[ \sin(kz), \cos(kz), 0 \right],
\ee
\be
\b = u_0 \sin(V_A kt) \left[ \cos(kz), -\sin(kz), 0 \right].
\ee
He then calculated an $\alpha$ effect from such helical waves and found
\be
\alpha(k) = - \frac{\gamma k < v^2 > }{(k{\overline B})^2 + \gamma^2}.
\ee
Here, $1/\gamma$ is the typical nonlinear damping time.\footnote{Note
that $\gamma$ must come from the nonlinear interactions between Alfv\'{e}n
wave modes of the type (B8) and (B9) {\it at different wave
lengths}. This is because, given the single mode standing wave pattern
like (B8) and (B9), the nonlinear terms in momentum equations, $\v
\times (\nabla \times \v)$ and $(\nabla \times \b) \times \b$, are both
zero, so that velocity field can be damped only through molecular
viscosity $\nu$, and the magnetic field can only through magnetic
viscosity, $\la$.} In the limit of large ${\overline B}$, (B10) can be
approximated as
\be
\alpha (k) \sim -\frac{1}{2{\overline B}^2 \tau} \left( <{\bf u} \cdot \v> -
<{\bf a} \cdot \b> \right),
\ee
where $\nabla \times {\bf u} = \v$ and $\nabla \times {\bf a} =
\b$. $\tau = 1/\gamma$ is the de-correlation time. This result is the
same as (59) of our calculation in large $\overline B$ limit, except
the following three aspects: first, in our model the de-correlation times
for $\v$ and $\b$ are different; second, the kinetic helicity and
current helicity in our model come from $\v^{(0)}$ and $\b^{(0)}$;
third, the nonlinear damping effect in our model comes from not only the
nonlinear interactions between modes of Alfv\'{e}n waves as proposed by
Kraichnan, but also the nonlinear interactions between $\{\v^{(0)}$,
$\b^{(0)}\}$. The standing helical waves in (B8) and (B9) are solution of
linear MHD equations with special initial conditions. In our model,
kinetic helicity and current helicity are determined by $\{\v^{(0)}$,
$\b^{(0)}\}$, which are driven by external helical forces. In many
rotating astrophysical systems, such external helical forces can be
buoyancy forces, or Coriolis forces. We believe the helical properties
of the linear Alfv\'{e}n wave of Kraichnan (1979), i.e., relations (B8) and
(B9), are determined by the MHD turbulence $\{\v^{(0)}$, $\b^{(0)}\}$,
which is in turn driven by external helical forces (see also Moffat,
1978). Such external energy sources to the MHD turbulence is of vast
importance to the dynamo action (Kulsrud, 1999), but have been
overlooked in previous calculations of $\alpha$ dynamo. Our work is
certainly an extension toward that direction.

\clearpage

\begin{deluxetable}{cc}
\footnotesize
\tablecaption{Notations and Their Physical Meanings}
\tablewidth{0pt}
\tablehead{ \colhead{Notation} & \colhead{Physical Meaning} }
\startdata
$\nu, \la$   & molecular viscosity and magnetic diffusivity \nl
$R_e, R_m$     & kinetic and magnetic Reynolds numbers \nl
$Pr=\nu/\la$  & magnetic Prandtl number \nl
$\alpha$     & $\alpha$ coefficient of dynamo theory \nl
$\OB$        & large-scale magnetic field, assumed constant \nl
$\v^{(0)}, \b^{(0)}$   & turbulent velocity and magnetic field that are
independent of $\OB$ \nl
$\v^{\prime}, \b^{\prime}$   & turbulent velocity and magnetic field
that depend on $\OB$ \nl
$l_0 = 1/k_0$        & size of energy containing eddies of the turbulence \nl
$k$          & magnitude of wave vector $|\bf k |$ \nl
$\omega_k$($\chi_k$)  & de-correlation frequency of velocity field $\v^{(0)}$($\b^{(0)}$) at $k$
\nl
$\tau_v$($\tau_b$)     & de-correlation time of turbulent velocity field $\v^{(0)}$($\b^{(0)}$) \nl
$R_{e,\omega}(k)$ & $={\omega_k}/{\nu k^2}, \sim (0.2 - 0.3)R_e$ at $k_0$ \nl
$R_{e,\chi}(k)$ & $={\chi_k}/{\nu k^2}, \sim (0.2 - 0.3)R_e$ at $k_0$ \nl
$R_{m,\omega}(k)$ & $={\omega_k}/{\la k^2}, \sim (0.2 - 0.3)R_m$ at $k_0$ \nl
$R_{m,\chi}(k)$ & $={\chi_k}/{\la k^2}, \sim (0.2 - 0.3)R_m$ at $k_0$ \nl
$\beta_1(k)$ & $=|\OB|/{\left( \omega_k/k \right) }$ \nl
$\beta_2(k)$ & $=|\OB|/{\left( \chi_k/k \right) }$ \nl
$\Delta_v = \int \Delta_{v}(k) dk$      & $\propto \tau_v <\v^{(0)} \cdot \nabla \times \v^{(0)}>$  \nl
$\Delta_b = \int \Delta_{b}(k) dk$     & $\propto \tau_b <\b^{(0)} \cdot \nabla \times \b^{(0)}>$  \nl
\enddata
\end{deluxetable}

\clearpage

\begin{figure}
\plotone{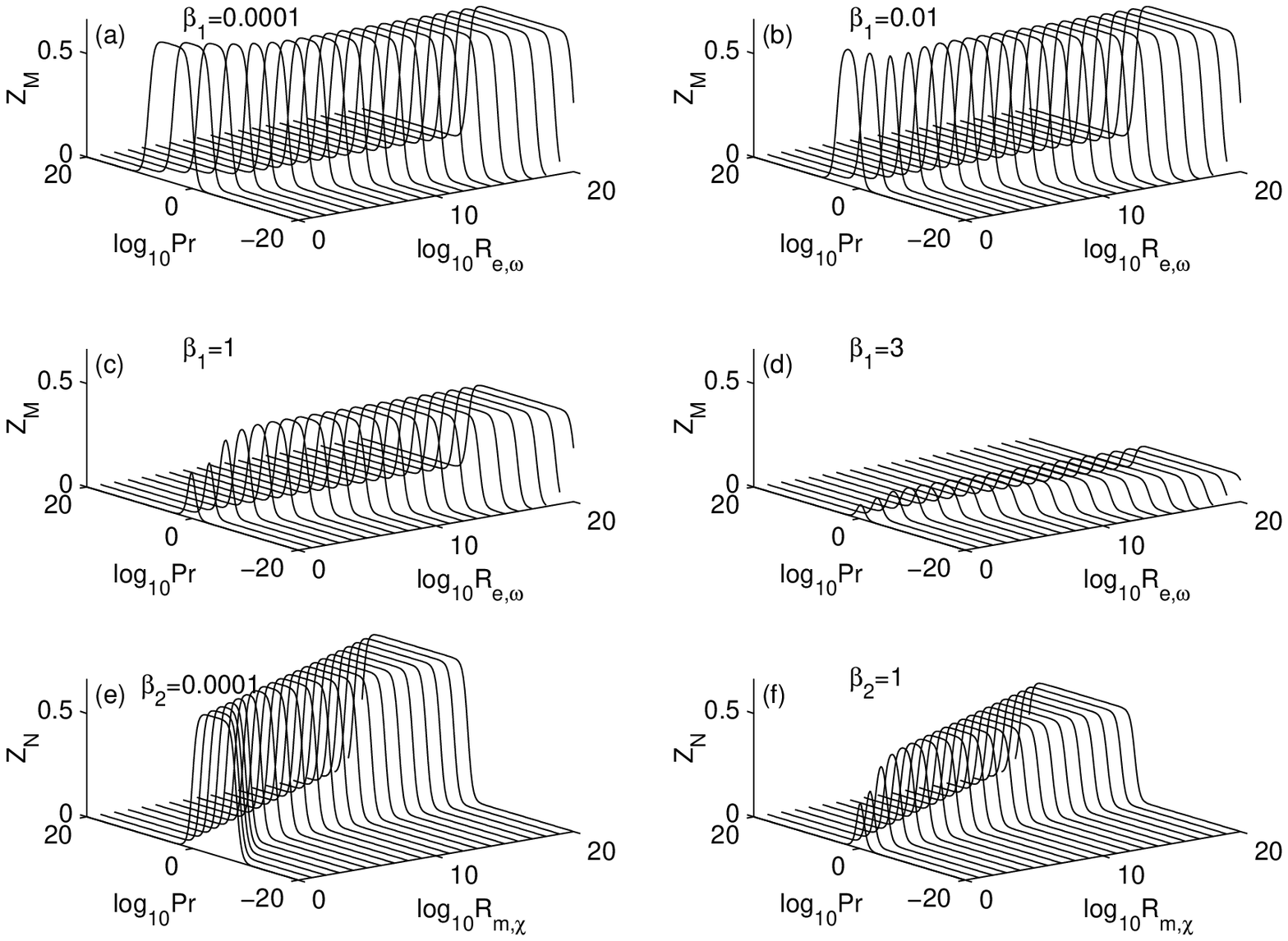}
\caption{This plot shows $Z_M$ and $Z_N$, defined in (75) and (76),
as functions of Prandtl number, kinetic Reynolds number and magnetic Reynolds
number at various values of $\beta_1$ and $\beta_2$ defined in (70) and
(71). 
}
\end{figure}
\begin{figure}
\plotone{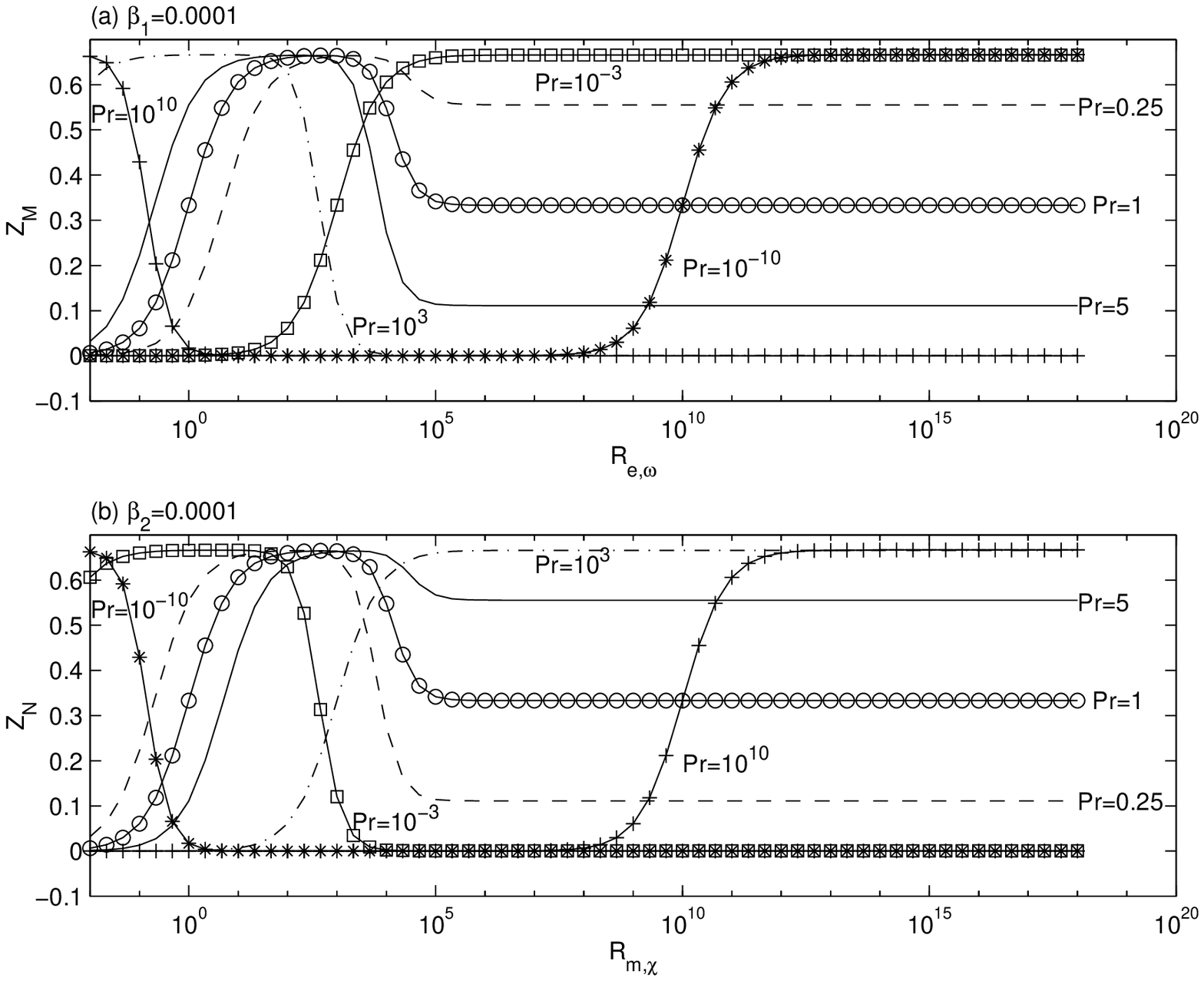}
\caption{$Z_M$(a) and $Z_N$(b) at various values of Prandtl
number. Here $\beta_1=\beta_2=0.0001$.}
\end{figure}
\begin{figure}
\plotone{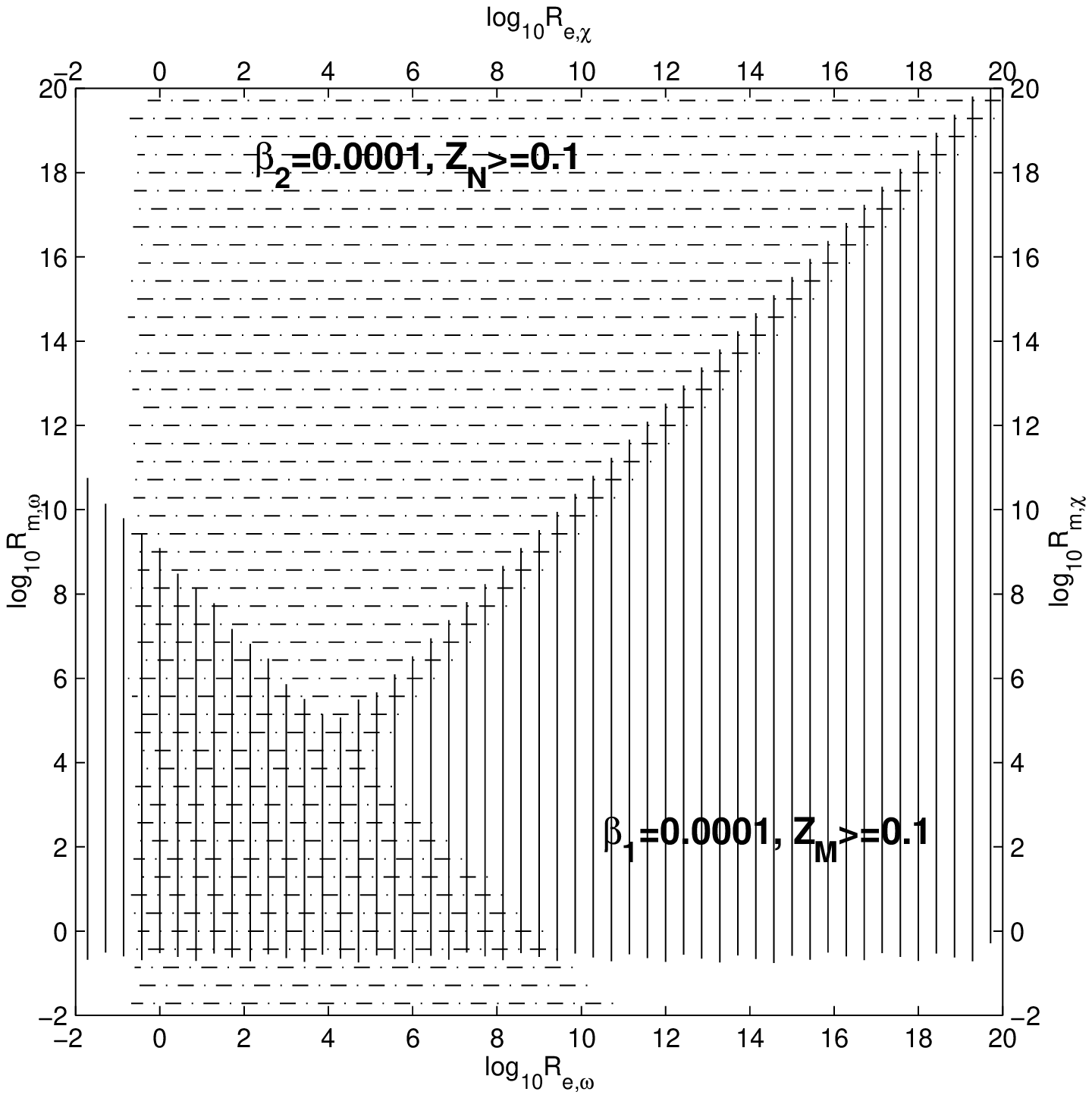}
\caption{The regimes where $Z_M \geq 0.1$(shaded by solid lines) or
$Z_N \geq 0.1$(shaded by dash-dot lines) at various values of kinetic and
magnetic Reynolds numbers. Here $\beta_1=\beta_2=0.0001$.}
\end{figure}
\begin{figure}
\plotone{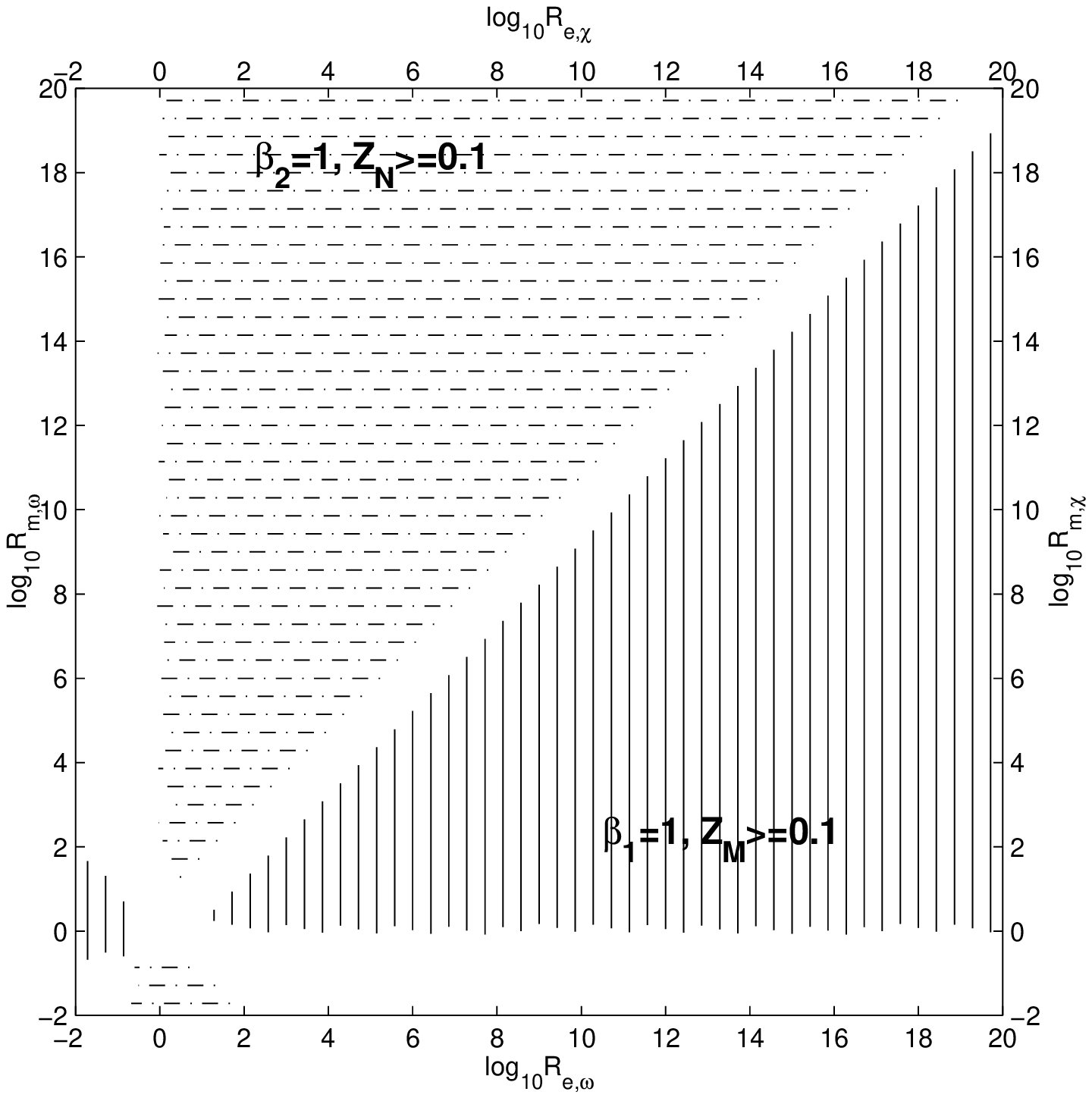}
\caption{The regimes where $Z_M \geq 0.1$(shaded by solid lines) or
$Z_N \geq 0.1$(shaded by dash-dot lines) at various values of kinetic and
magnetic Reynolds numbers. Here $\beta_1=\beta_2=1$. Notice that the
area covered by the shade is smaller than the shaded area shown in
Figure 3. This is due to the quenching of $\alpha$-effect by large
value of ${\overline B}$.}
\end{figure}

\end{document}